\pdfoutput=1

\documentclass[11pt,twoside,a4paper,cmspaper,final,collab]{cms-tdr}
\def\svnVersion{296236}\def\cmsCernNo{2013-037}\def\cmsCernDate{\today}\def\cmsMessage{Published in the European Physical Journal C as \href{http://dx.doi.org/10.1140/epjc/s10052-015-3491-9}{\doi{10.1140/epjc/s10052-015-3491-9.}}}

\begin{document}\cmsNoteHeader{QCD-11-006}

\hyphenation{had-ron-i-za-tion}
\hyphenation{cal-or-i-me-ter}
\hyphenation{de-vices}
\RCS$Revision: 296202 $
\RCS$HeadURL: svn+ssh://svn.cern.ch/reps/tdr2/papers/QCD-11-006/trunk/QCD-11-006.tex $
\RCS$Id: QCD-11-006.tex 296202 2015-07-12 16:02:48Z banerjee $
\newlength\cmsFigWidth
\ifthenelse{\boolean{cms@external}}{\setlength\cmsFigWidth{0.98\columnwidth}}{\setlength\cmsFigWidth{0.485\textwidth}}
\newlength\cmsFigWidthA
\ifthenelse{\boolean{cms@external}}{\setlength\cmsFigWidthA{0.98\columnwidth}}{\setlength\cmsFigWidthA{0.39\textwidth}}
\ifthenelse{\boolean{cms@external}}{\providecommand{\cmsLeft}{top\xspace}}{\providecommand{\cmsLeft}{left\xspace}}
\ifthenelse{\boolean{cms@external}}{\providecommand{\cmsRight}{bottom\xspace}}{\providecommand{\cmsRight}{right\xspace}}
\providecommand{\as}{\alpha_{\mathrm{s}}}
\providecommand{\lumInt}{\ensuremath{\int{\mathcal{L}}}\xspace}
\providecommand{\MADGRAPHPYTHIASIX}{\MADGRAPH{+}\PYTHIA{6}\xspace}
\newcommand{\cbz}{\ensuremath{\chi_{\mathrm{BZ}}}\xspace}
\newcommand{\cnr}{\ensuremath{\theta_{\mathrm{NR}}}\xspace}
\title{Distributions of topological observables in inclusive three- and four-jet events in pp collisions at $\sqrt{s} = 7$\TeV}

\date{\today}

\abstract{
This paper presents distributions of topological observables in inclusive three- and four-jet events produced in pp collisions at a centre-of-mass energy of 7\TeV with a data sample collected by the CMS experiment corresponding to a luminosity of 5.1\fbinv.
The distributions are corrected for detector effects, and compared with several event generators based on two- and multi-parton matrix elements at leading order. 
Among the considered calculations,  \MADGRAPH interfaced with \PYTHIA{6} displays the overall best agreement with data.
}

\hypersetup{
pdfauthor={CMS Collaboration},%
pdftitle={Distributions of topological observables in inclusive three- and four-jet events in pp collisions at sqrt(s) = 7 TeV},%
pdfsubject={CMS},%
pdfkeywords={CMS, QCD}}

\maketitle

\section{Introduction}\label{sec:intro}

{\tolerance=1500
In proton-proton collisions at the LHC, interactions
take place between the partons of the colliding protons.
The scattered partons from hard collisions fragment and hadronize into
collimated groups of particles called jets.
The study of jets with high transverse momentum (\pt) provides a test of
the predictions from quantum chromodynamics (QCD) and deviations from
these predictions can be used to look for physics beyond the standard model.
While parton scattering is an elementary QCD process that can be calculated
from first principles,  predictions of jet distributions require an accurate
hadronization model. In this paper, several hadronization models are examined.
\par}

High-\pt parton production is described by perturbative QCD (pQCD) in terms
of the scattering cross section convolved with a parton distribution function
(PDF) for each parton that parametrizes the momentum distribution of
partons within the proton. The hard-scattering cross section itself can be
written as an expansion in the strong coupling constant $\as$. The leading
term in this expansion corresponds to the emission of two partons. The next term
includes diagrams where an additional parton is present in the final state as a
result of hard-gluon radiation (\eg $\Pg\Pg\to\Pg\Pg\Pg$). Cross sections for
such processes diverge when any of the three partons becomes soft or when two
of the partons become collinear. Finally, pQCD predicts three classes of
four-jet events that correspond to the processes $\PQq\PQq/\Pg\Pg \to \PQq\PQq\Pg\Pg, \PQq\PQq/\Pg\Pg
\to \PQq\PQq\PQq\PQq$ and $\PQq\Pg \to \PQq\Pg\Pg\Pg/\PQq\PQq\PQq\Pg$, where $\PQq$ stands for both
quarks and anti-quarks. Processes with two or more gluons in the final state
receive a contribution from the triple-gluon vertex, a consequence of the
non-Abelian structure of QCD.

We are studying distributions of topological variables, which are sensitive
to QCD color factors, the spin structure of gluons, and hadronization models.
These topological variables were studied widely in the earlier
LEP \cite{lep1,lep2} and the Tevatron \cite{tev1,tev2}
experiments and help to validate theoretical models implemented in various
Monte Carlo (MC) event generators.

The distributions of multijet variables are sensitive to the treatment of
the higher-order processes and approximations involved.
Many MC event generators make use of leading order (LO) matrix
elements (ME) in the primary 2$\to$2 process.
A good agreement between the measurements and MC
predictions can establish the validity of the treatment of
higher-order effects, and any large deviation may lead to large
systematic uncertainties in searches for new physics.

The multijet observables presented here are based on hadronic
events from 7\TeV pp collision data recorded with the CMS detector
corresponding to an integrated luminosity of 5.1\fbinv. The
kinematic and angular properties of these events are computed from the jet
momentum four-vectors. Unfolding techniques are used to correct for the
effects of the detector resolution and efficiency. Systematic uncertainties
resulting from the limited knowledge of the jet energy scale (JES),
jet energy and angular resolution (JER), unfolding, and
event selection are estimated, and the unfolded distributions are compared with
predictions of several QCD-based MC models.

In this paper, the CMS detector is briefly described in Section
\ref{sec:cmsdet}. Sections \ref{sec:mcmodels} and \ref{sec:defs} summarize the
MC models used and the variables studied in this paper. Event
selection and measurements are described in Sections \ref{sec:data} and
\ref{sec:measure}, respectively.  The correction of the distributions due to
detector effects is discussed in Section \ref {sec:unfold}. Sections
\ref{sec:syst} and \ref{sec:results} describe the estimation of systematic
uncertainties and the final results. The overall summary is given in Section
\ref{sec:summ}.

\section{The CMS detector}\label{sec:cmsdet}

The central feature of the CMS apparatus is a superconducting solenoid of
6\unit{m} internal diameter, providing a magnetic field of 3.8\unit{T}. Within
the field volume are a silicon pixel and strip tracker, a
lead tungstate crystal electromagnetic calorimeter (ECAL), and a brass
and scintillator hadron calorimeter (HCAL), each composed of a barrel and
two endcap sections. Muons are measured in gas-ionization detectors embedded
in the steel flux-return yoke outside the solenoid. Extensive forward
calorimetry complements the coverage provided by the barrel and endcap
detectors. The barrel and endcap calorimeters cover a pseudorapidity region
$-3.0<\eta<3.0$. Pseudorapidity is defined as $\eta = - \ln\tan[\theta/2]$, where $\theta$ is the polar angle. The transition between barrel and endcaps happens
at $\abs{\eta} = 1.479$ for the ECAL and $\abs{\eta}= 1.15$ for the HCAL.
The first level (L1) of the CMS trigger system, composed of custom hardware
processors, uses information from the calorimeters and muon detectors to
select the most interesting events in a fixed time interval of less than 4\mus.
The high-level trigger (HLT) processor farm further decreases the event rate
from around 100\unit{kHz} to around 400\unit{Hz} before data storage.
A more detailed description of the CMS detector, together with a definition of
the coordinate system used and the relevant kinematic variables, can be found in
Ref.~\cite{cmsdet}.

\section{Monte Carlo models}\label{sec:mcmodels}

The MC event generators rely on models using modified LO QCD
calculations. The elementary hard process between the partons is computed
at LO. The parton shower (PS), used to simulate higher-order processes,
follows an ordering principle motivated by QCD. Nevertheless, the parton shower
models can differ in the ordering of emissions and the event generators can
also have different treatments of beam remnants and multiple interactions.

The \PYTHIA 6.4.26 \cite{pythia6} event generator uses a PS model to
simulate higher-order processes \cite{hopr1,hopr2,hopr3} after the LO ME
from pQCD calculations. The PS model, ordered by the \pt of the emissions,
provides a good description of event shapes when the emitted partons
are close in phase space. Events are generated with the Z2 tune \cite{z2tune}
for the underlying event. This tune is identical to the Z1 tune
\cite{z1tune}, except that it uses CTEQ6L1 \cite{cteq} PDFs.
The partons are hadronized (process of converting the partons into measured
particles) using the Lund string model \cite{lund1,lund2}.

The \PYTHIA 8.153~\cite{pythia8} event generator also uses a PS model
with the successive emissions of partons ordered in \pt and the Lund string model for hadronization. The main difference between the
two \PYTHIA versions is the description of multiparton
interactions (MPI).  In \PYTHIA{8}, initial state radiation (ISR), final
state radiation (FSR), and MPI are interleaved in the \pt ordering, while in
\PYTHIA{6}, only ISR and FSR are interleaved.
The \textsc{tune4C} \cite{pythia8tune2c} is used with this generator.
This tune uses CTEQ6L1 PDFs with parameters using CDF as well
as early LHC measurements.

The \HERWIGpp 2.4.2 \cite{herwig}  \textsc{tune23} \cite{tune23}
program takes the LO ME and simulates a PS using the
coherent branching algorithm with angular ordering \cite{angularorder} of the
showers. The partons are hadronized in this model using a cluster model
\cite{cluster} and the underlying event is simulated using the eikonal multiple
partonic scattering model.

In the case of \MADGRAPH 5.1.5.7~\cite{madgraph}, multiparton final
states are also computed at tree level. The parton shower and
nonperturbative parts for \textsc{Madgraph 5.1.5.7} simulation sample is
handled by \PYTHIA 6.4.26 with Z2 tune. The MLM matching
procedure \cite{mlm} is used to avoid double counting between the ME and
the PS. The \MADGRAPH samples are created in four
bins of the variable $\HT$, the scalar sum of the parton \pt{}s. The
matching between ME and PS has been studied in detail and has been
validated using inclusive jet \pt distributions. Several samples are
generated using different matching parameters and are used in estimating
systematic uncertainty in the theoretical prediction.

These MC programs are the most commonly used models to describe multi-partonic
final states and are normally used to describe QCD background in searches
within CMS. The events produced from these models are simulated using a CMS
detector simulation program based on \GEANTfour \cite{g4} and reconstructed
with the same program used for the data. These MC events are used for the
comparison with the measurements as well as to correct the distributions for
detector effects.

\section{Definition of variables}\label{sec:defs}
\subsection{Three-jet variables}
The topological variables used in this study are defined in the parton or jet
centre-of-mass (CM) system. The topological properties of the three-parton
final state in the CM system can be described in terms of five
variables \cite{tev1}. Three of the variables reflect partition of the
CM energy among the three final-state partons. There are three angles, which
define the spatial orientation in the plane containing the three partons,
but only two are independent.

\begin{figure}[htbp]
\centering
\includegraphics[width=\cmsFigWidth]{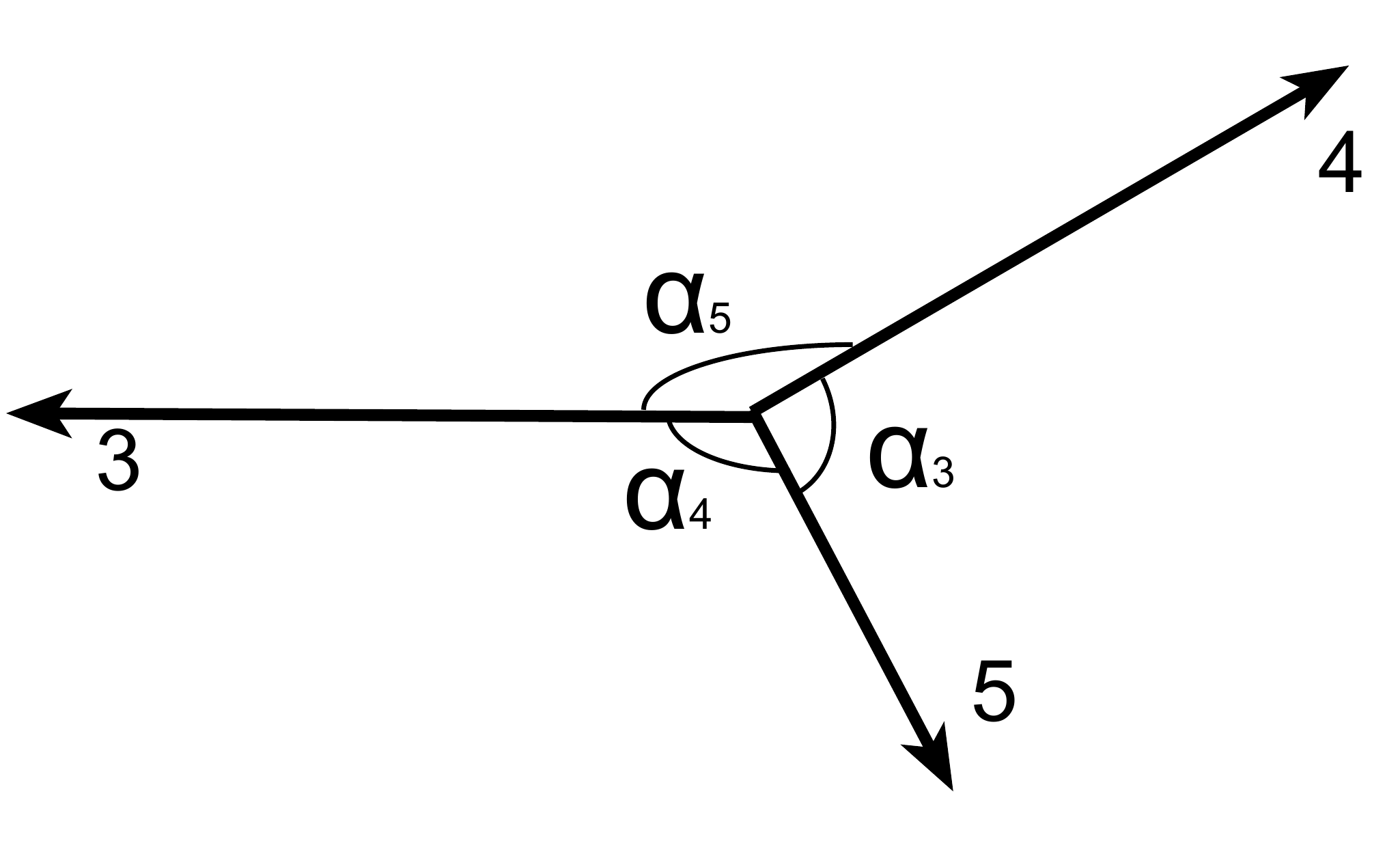}

\caption{Illustration of the three-jet variables in the process 1+2$\to$3+4+5.
The scaled energies are related to
         the angles ($\alpha_i$) among the jets for massless parton.}
\label{fig:psitheta}
\end{figure}

It is convenient to introduce the notation 1+2$\to$3+4+5 for
the three-parton process. Here, numbers 1 and 2 refer to incoming partons while
the numbers 3, 4, and 5 label the outgoing partons in a descending order in
energies in the parton CM frame, \ie $E_3>E_4>E_5$
(Fig.~\ref{fig:psitheta}).
The final-state parton energy is an obvious choice for the topological
variable for the three-parton final state.  For simplicity, $E_i$ ($i =
3$, 4, 5) is often replaced by the scaled variable $x_i$ ($i = 3$, 4, 5), which
is defined by $x_i = 2E_i/\sqrt{\hat{s}_{345}}$, where $\sqrt{\hat{s}_{345}}$
is the CM energy of the hard-scattering process. It is also
referred to as the mass of the three-parton system, and by definition,
\begin{equation}
\label{eqn:conserv}
x_3 + x_4 + x_5  =  2.
\end{equation}

The internal structure of the three-parton final state is determined by any
two scaled parton energies. The third one is calculated using Equation
\ref{eqn:conserv}. It needs two angular variables which fix the event
orientation.
In total, five independent kinematic variables are needed to describe the
topological properties of the three-parton final state. In this analysis,
however, the study is restricted to three variables: $\sqrt{\smash[b]{\hat{s}_{345}}}$,
$x_3$, and $x_4$, while the angular variables are not included.

\subsection{Four-jet variables}
To define a four-parton final state in its CM frame, eight
independent parameters are needed. Two of these define the overall event
orientation, while the other six fix the internal structure of the four-parton
system. In contrast to the three-parton final state, there is no simple
relationship between the scaled parton energies and the opening angles between
partons. Consequently, the choice of topological variables is less obvious in
this case. Variables are defined here in a way similar to those investigated
for the three-parton final state. The four partons are ordered in descending
energy in the parton CM frame and labeled from 3 to 6. The
variables include the scaled energies and the polar angles of the four partons
with respect to the beams.

\begin{figure}[htbp]
\centering
 \includegraphics[width=\cmsFigWidth]{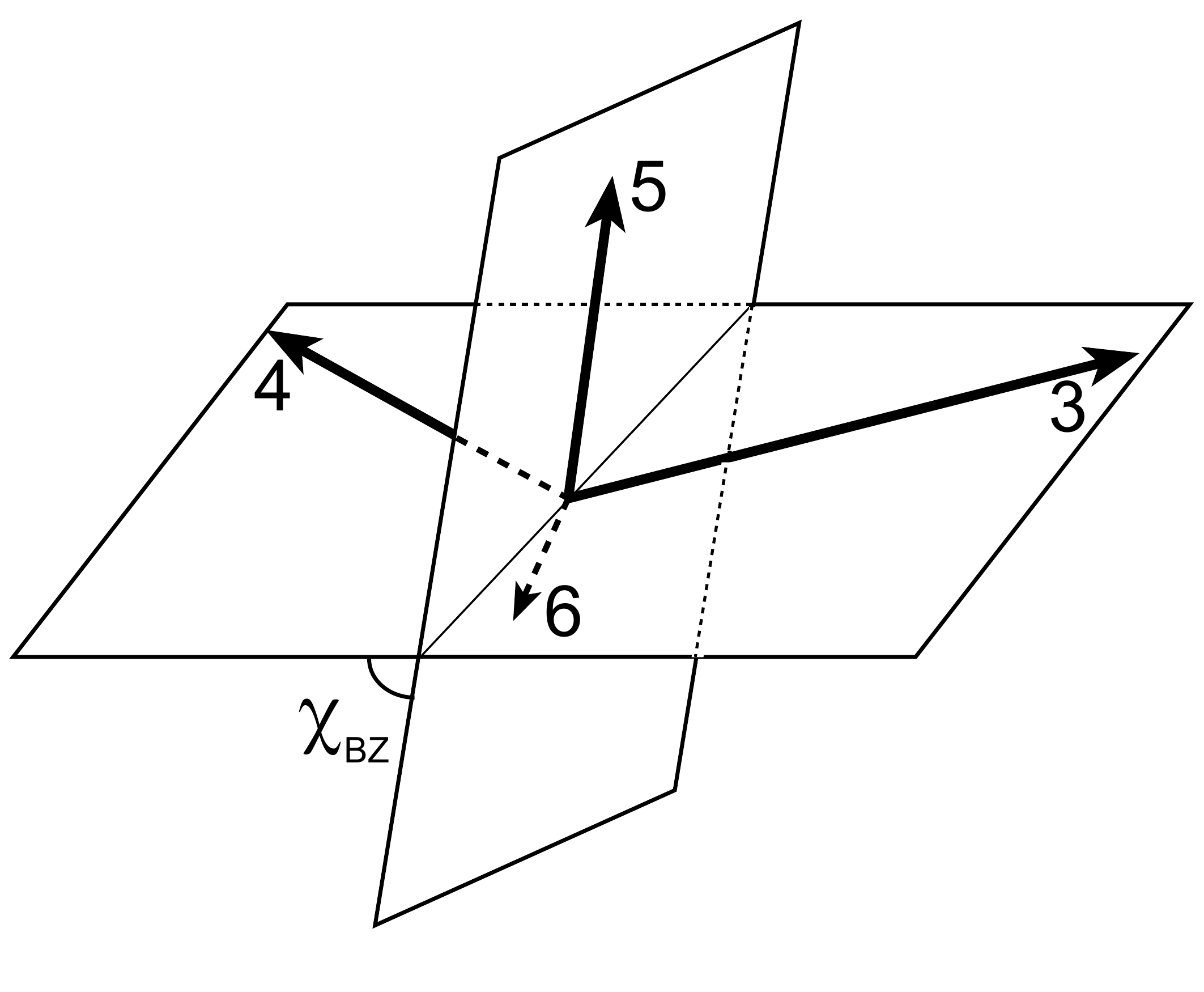}
 \includegraphics[width=\cmsFigWidth]{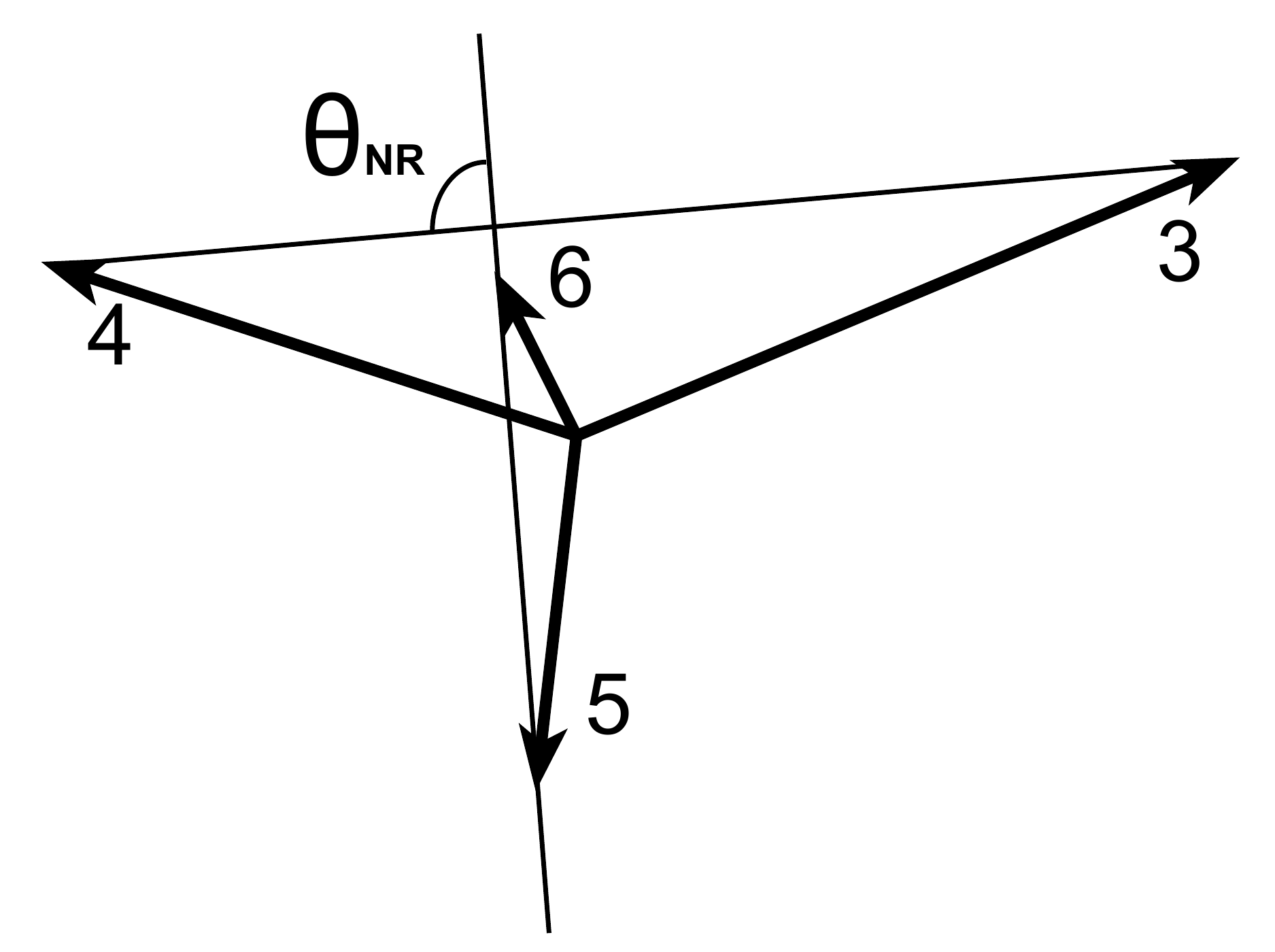}

\caption{Illustration of the Bengtsson--Zerwas angle ($\cbz$) and the
         Nachtmann--Reiter angle ($\cnr$) definitions for the four-jet
         events. The \cmsLeft figure shows the Bengtsson--Zerwas angle, which is the
         angle between the plane containing the two leading jets and the plane
         containing the two nonleading jets. The \cmsRight figure shows the
         Nachtmann--Reiter angle, which is the angle between the momentum vector
         differences of the two leading jets and the two nonleading jets.}
\label{fig:bznr}
\end{figure}

In addition to the  four-parton CM energy or the mass of the four-parton
system ($\sqrt{\hat{s}_{3456}}$), two angular distributions characterizing the
orientation of event planes are investigated. One of these is the
Bengtsson--Zerwas angle ($\cbz$) \cite{bz} defined as the angle between the
plane containing the two leading jets and the plane containing the two
nonleading jets:
\begin{equation}
\cos\cbz  =  \frac{(\vec{p}_3\times\vec{p}_4)\cdot(\vec{p}_5\times\vec{p}_6)}
                    {\abs{\vec{p}_3\times\vec{p}_4}\abs{\vec{p}_5\times\vec{p}_6}}.
\end{equation}
The second variable is the cosine of the Nachtmann--Reiter angle
($\cos\cnr$) \cite{nr}  defined as the angle between the momentum vector
differences of the two leading jets and the two nonleading jets:
\begin{equation}
\cos\cnr  =  \frac{(\vec{p}_3 - \vec{p}_4)\cdot(\vec{p}_5 - \vec{p}_6)}
                    {\abs{\vec{p}_3 - \vec{p}_4}\abs{\vec{p}_5 - \vec{p}_6}}.
\end{equation}
Figure \ref{fig:bznr} illustrates the definitions of $\cbz$ and
$\cnr$ variables. Historically, $\cbz$ and $\cnr$ were
proposed for $\Pep\Pem$ collisions to study gluon self-coupling. Their
interpretation in pp collisions is more complicated, but the variables can be
used as a tool for studying the internal structure of the four-jet events.

\section{Data samples and event selection}\label{sec:data}

Jets are reconstructed from particle-flow (PF)
objects \cite{pflow,pflow1} using the anti-\kt clustering algorithm
\cite{jetalgo} with the distance parameter $R = 0.5$, as calculated with
\textsc{Fastjet} 2.0 \cite{fastjet}. The PF algorithm utilizes the
best energy measurements of each particle candidate from the most suitable
combination of the detector components.
A cluster is formed from all the particle-flow candidates that satisfy the
chosen distance parameter. The four-momentum of the jet is then defined as
the sum of four-momenta of the corresponding particle-flow candidates, which
results in jets with nonzero mass.

The JES correction applied to jets used in this analysis is based on
high-\pt jet events generated by \PYTHIA{6} and then simulated
using \textsc{Geant4}, and in situ measurements with dijet and
photon+jet events \cite{jec}.
An average of ten minimum bias interactions occur in each pp bunch crossing
(pileup), and this requires an additional correction to remove the extra energy
deposited by these pileup events.
The size of the correction depends on the \pt and $\eta$ of the jet.
The correction appears as a
multiplicative factor to the jet energy, and is typically less than 1.2 and
approximately uniform in $\eta$.

Events passing single-jet HLT requirements are used in this analysis. These
triggers require jets reconstructed from calorimetric information with the
anti-$\kt$ clustering algorithm and with energy corrections applied.
Jets are ordered in decreasing jet \pt, and
the leading jet \pt is required to be above a certain threshold.
As offline jets are reconstructed with the PF algorithm, this may
result in a trigger not being fully efficient near the threshold.
Trigger efficiencies are studied as a function of the leading jet \pt for
all trigger thresholds. Values of the leading jet \pt, where the trigger
efficiency is determined to be larger than 99\%, are listed in Table~\ref{tab:trigger}. It also summarizes the prescale factors and the effective
integrated luminosities collected using the different HLT thresholds.

\begin{table*}[htbp]
\topcaption{Prescales, integrated luminosity and offline \pt threshold of
           the leading jet for different trigger paths. The terminology for
           Level 1 (L1) triggers as well as HLT includes the jet \pt            threshold (in \GeV) applicable to the trigger.}
\centering\begin{tabular}{ccccccc}\hline
Period & HLT &  HLT60     &  HLT110    &  HLT190    &  HLT240 & HLT370 \\
       & L1  &SingleJet36 &SingleJet68 &SingleJet92 &SingleJet92&SingleJet92/ \\
       &     &            &            &            &           &SingleJet128 \\
\hline
2011A  & L1 prescale    & 1--300  & 1--10   &  1     &  1     &  1   \\
       & HLT prescale   & 15--180 & 1--5000 &  1--60 &  1--24 &  1   \\
       & $\lumInt$ (\pbinv)& 0.29 & 6.16 & 114.7 & 392.2 & 2328 \\ \hline
2011B  & L1 prescale    & 50--400 & 1--20    & 1--10  &  1    &  1   \\
       & HLT prescale   & 80--84  & 80--1000 & 10--100& 4--30 &  1   \\
       & $\lumInt$ (\pbinv) & 0.12 & 1.12 &  40.2 & 136.0 & 2767 \\ \hline
Overall & $\lumInt$ (\pbinv)& 0.41 & 7.29 & 154.8 & 528.2 & 5096 \\ \hline
        & \pt threshold & 110\GeV & 190\GeV& 300\GeV & 360\GeV & 500\GeV\\
\hline\end{tabular}
\label{tab:trigger}
\end{table*}

Jets are selected with restrictive criteria on the neutral energy
fractions (both electromagnetic and hadronic components), and all the jets
are required to have $\pt > 50$\GeV and absolute rapidity, ($y =
(1/2)\ln[(E+p_z)/(E-p_z)]$), $\abs{y}\leq 2.5$. The jet with the
highest \pt is required to be above a threshold as given by the requirement
from the trigger turn-on curve. To avoid overlap of events from two different
HLT paths, the \pt of the leading jet is also required to be less than an
upper value. The overall criteria are summarized in Table \ref{tab:jetlim}.
Though data from all the five trigger paths are studied, figures from two
representative trigger paths (the highest \pt threshold and a lower one
with good statistical accuracy) are presented in this paper.

\begin{table*}[htbp]
\topcaption{Threshold of the leading jet \pt for different HLT paths. This
           paper shows results from two representative trigger paths HLT110 and
           HLT370.}
\centering\begin{tabular}{cccccc}\hline
 HLT &  HLT60     &  HLT110    &  HLT190    &  HLT240 & HLT370 \\ \hline
 Leading jet \pt (\GeVns{}) & 110--190 & 190--300 & 300--360 & 360--500 & $>$500 \\ \hline
\end{tabular}
\label{tab:jetlim}
\end{table*}

Events are selected with at least three jets passing the selection criteria as
stated above. Additional selection requirements are also applied to reduce
backgrounds due to beam halo, cosmic rays and detector noise. The event must
have at least one good reconstructed vertex~\cite{pv}. Missing transverse
energy, \ETm, is required to be less than $0.3 \sum \ET$,
where the summation is over all PF jets. The quantities \ETm and $\sum \ET$ are obtained from negative
vector sum and scalar sum of the transverse momenta of the jets, respectively.
A number of event filters \cite{noise} accept only those events that have
negligible noise in the detector.
The jets are ordered in decreasing \pt, and an event with at least three
(four) jets satisfying the jet selection criteria is classified as a three-jet
(four-jet) event.

\section{Measurements}\label{sec:measure}

The 4-momenta of all the jets in the three- or four-jet event category are
transformed into the CM frame of the three- or four-jet system. The
jets are then ordered in decreasing energy. The three- and four-jet variables
as described in Section \ref{sec:defs} are then calculated from the kinematic
and angular information of the jets.
Since detector resolution  varies over the potential kinematic ranges, variable
bin widths are adopted for the jet masses and the scaled jet energies, while for
angular variables constant bin widths are used.

\subsection{Detector-level distributions}

The measured distributions of the three- and four-jet variables are compared
with predictions from two MC generators (\PYTHIA{6} and
\MADGRAPHPYTHIASIX), simulated using the identical detector condition as that
in the data. The identical pileup condition is obtained by reweighting the MC
simulation to match the spectrum of pileup interactions observed in the
data. The size of the reweighting correction is typically less than 1\%.
The agreement between the data and the MC predictions is reasonable, so
these MC generators are used to correct the measured distributions.

\begin{figure}[h!tbp]
\centering
 \includegraphics[width=\cmsFigWidthA]{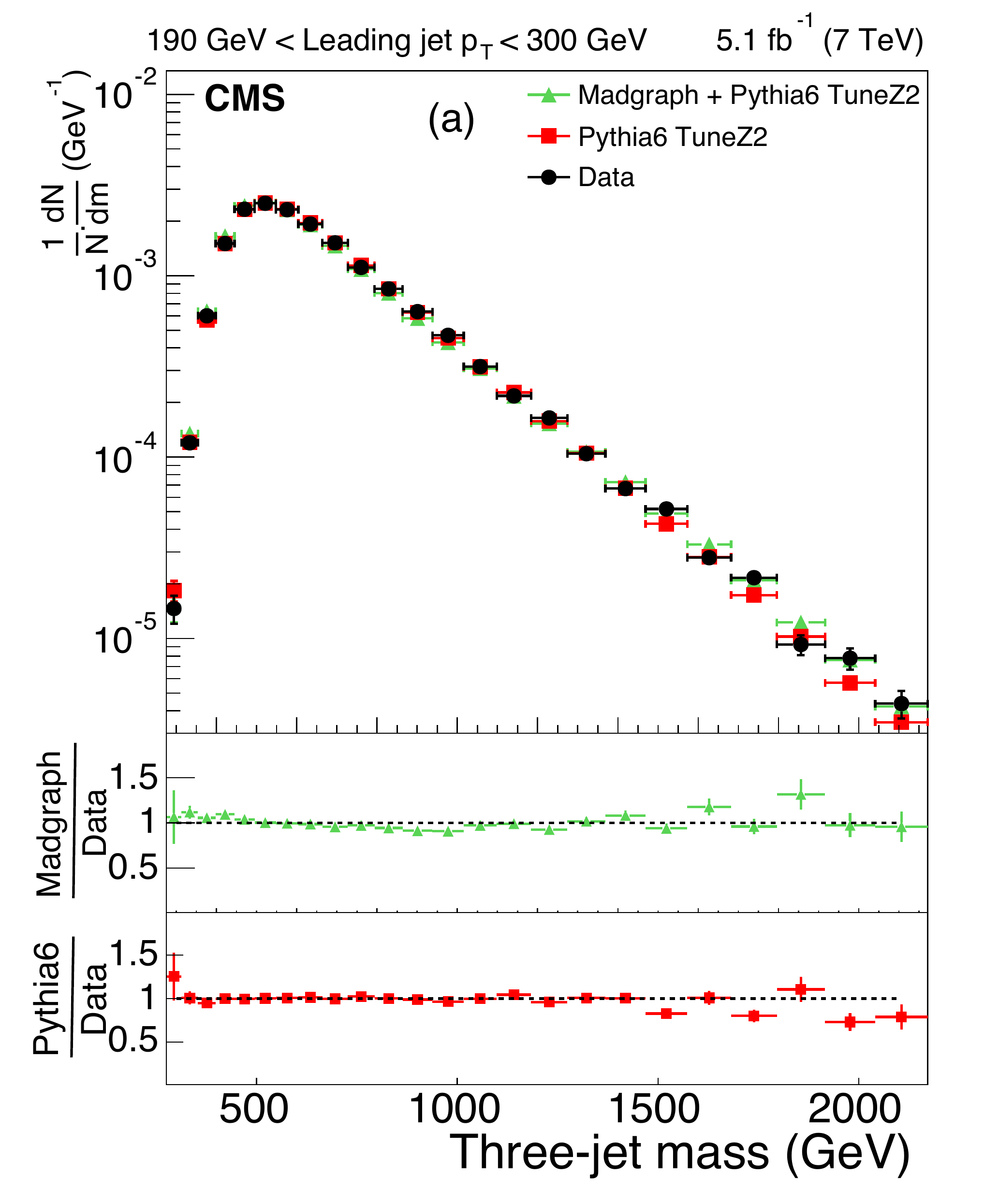}
 \includegraphics[width=\cmsFigWidthA]{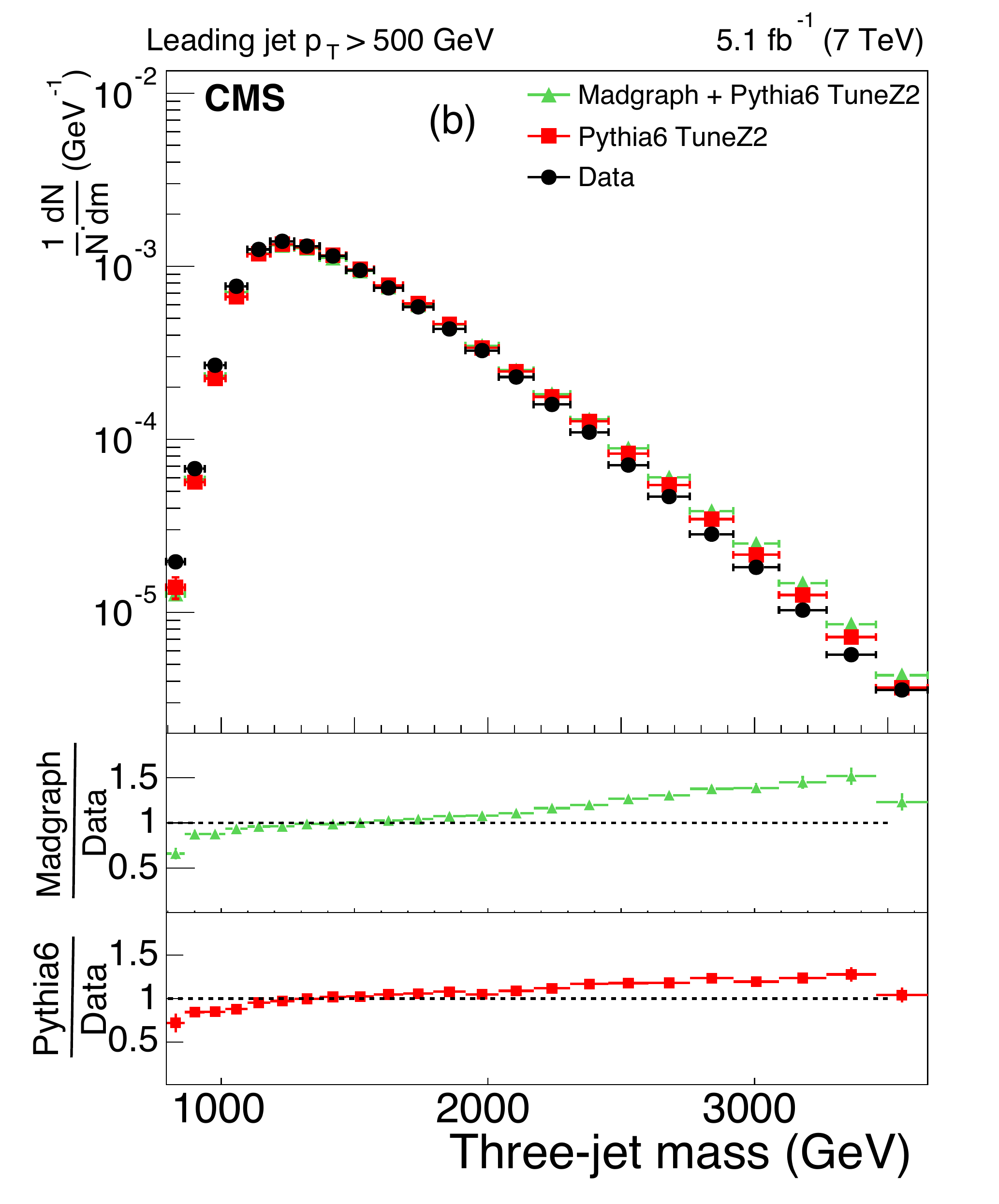}

\caption{The upper panels display the normalized distributions of the
         reconstructed three-jet mass for events where the most forward jet has
         $\abs{y}<2.5$. Figures differ by \pt ranges of the leading jet:
         190--300\GeV (a), and above 500\GeV (b) for data (before
         correction due to detector effects) and predictions from MC generators.
         The bottom panel of each plot shows the ratio of MC predictions to the
         data. The data are shown with only statistical uncertainty.}
\label{fig:m3jetmass}
\end{figure}

\begin{figure}[h!tbp]
\centering
 \includegraphics[width=\cmsFigWidthA]{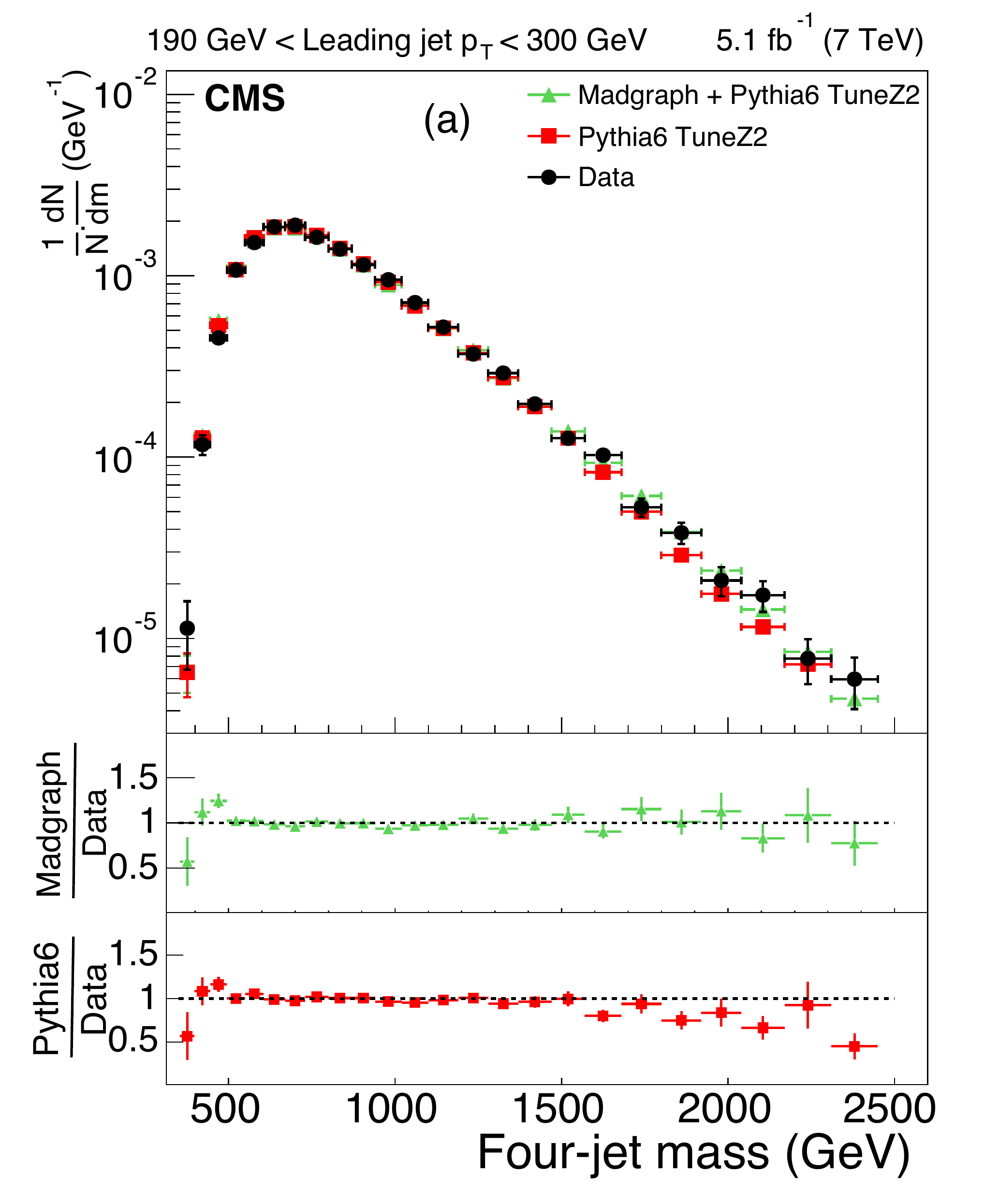}
 \includegraphics[width=\cmsFigWidthA]{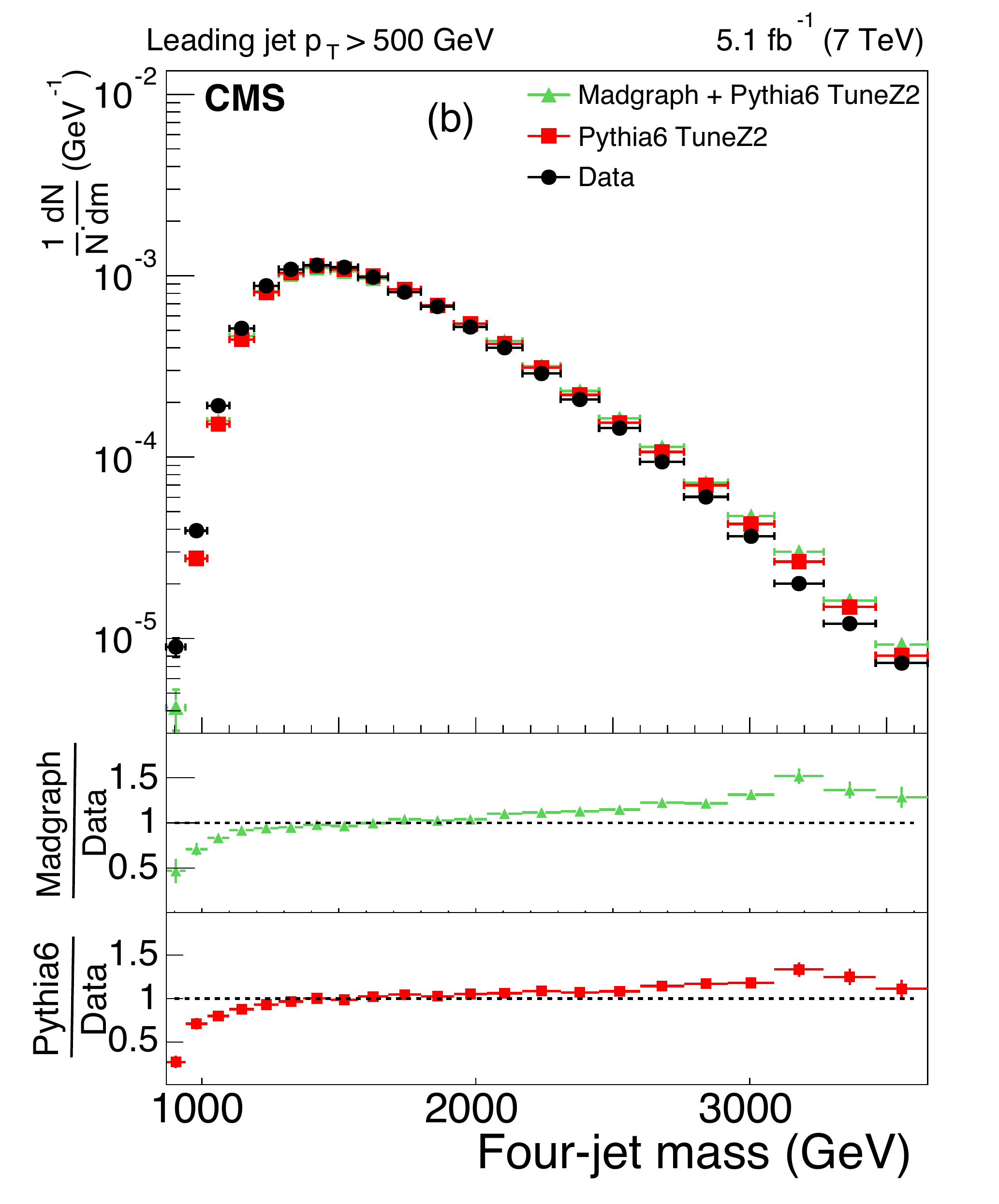}

\caption{The upper panels display the normalized distributions of the
         reconstructed four-jet mass for events where the most forward jet has
         $\abs{y}< 2.5$.  The other explanations are the same
         as Fig.~\ref{fig:m3jetmass}.}
\label{fig:m4jetmass}
\end{figure}

Figures \ref{fig:m3jetmass}  and \ref{fig:m4jetmass} show the normalized three-
and four-jet mass distributions.
The data are compared with two different MC programs: \PYTHIA{6}
and \MADGRAPHPYTHIASIX, each with two different HLTs
with \pt thresholds above 110 and 370\GeV.
As can be seen from the figures, there is
agreement within a few percent between the data and the predictions of these
two simulations. The difference between the predictions and the
data varies typically from 4\% to 10\%. However, there is a systematic
deviation observed at high masses where the simulations are higher than the
data.

\section{Corrections for detector effects}\label{sec:unfold}

Multijet variables obtained from MC samples may differ from data because of the
detector resolution and acceptance. Before comparisons with other experiments or
theoretical predictions can be made, detector effects are unfolded into
distributions at the final-state particle level.
The basic component of the unfolding is the response function, where
experimental observables are expressed as a function of theoretical
observables. For simplicity, observables are taken in discrete sets, and the
response function is replaced by a response matrix. The observed
distribution is then unfolded with the inverse of response matrix to obtain a
distribution corrected for detector effects. Matrix inversion has
potential complications, because it cannot handle large statistical
fluctuations and the matrix itself could be singular. Instead, we use
the RooUnfold package \cite{roounfold} with the
D'Agostini iterative method \cite{bayes} as the default algorithm and the
singular value decomposition method \cite{SVD} for cross-checks.

{\tolerance=400
The default response matrix is obtained using the \PYTHIA{6} event
generator.
Statistical uncertainties are estimated from the square root of the
covariance matrix obtained from a variation of the results generated by
simulated experiments.
\par}

The corrections for detector resolution and acceptance change the shape of
the three-jet mass distributions by approximately 10\%, less than 5\% for the
scaled energy of nonleading jets, and up to 20\% for the scaled energy of the
leading jet. For four-jet variables, corrections applied are of the order of
20\% for the four-jet mass, 10\% for $\cbz$, and less than 5\% for $\cos\cnr$.

\section{Systematic uncertainties}\label{sec:syst}

The leading sources of systematic uncertainty are due to the JES, the JER,
and the model dependence of corrections to the data.
The distributions are presented in this analysis as normalized distributions,
thus the absolute scale uncertainty of energy measurement does not play
a significant role. There are insignificant contributions
due to resolution of $y$. The main contribution of JES or JER
to the uncertainty in the measurements is due to the
migration of events from one category of jet multiplicity to the other.

The effect of pileup in the measured distributions has been studied
as a function of the number of reconstructed vertices in the event.
None of the variables show any significant dependence on the
pileup condition, so systematic uncertainty due to pileup can be neglected.

\subsection{Jet energy scale}

One of the dominant sources of systematic uncertainty is due to the jet energy
scale corrections. The JES uncertainty has been estimated to be 2--2.5\% for
PF jets~\cite{jec}, depending on the jet \pt and $\eta$. In order
to map this uncertainty to the multijet variables, all jets in the selected
events are systematically shifted by the respective uncertainties, and a new
set of values for the multijet variables is calculated. This causes a migration
of events from an event category of a given jet multiplicity to a different jet
multiplicity. The migration could be as high as 20\% for some of the event
categories. The corresponding distributions are then unfolded using the
standard procedure as described in Section \ref{sec:unfold}.  The difference of
these values from the central unfolded results is a measure of the uncertainty
owing to the JES.

Uncertainties owing to the JES are found to be between 0.2--5.5\% in the
three-jet mass, and 0.3--10\% in the four-jet mass. The systematic
uncertainties are the largest at both ends of the mass spectra. The systematic
uncertainties in scaled energy are between 0.1\% and 2.0\%, and those in
angular variables are in the range 0.1--3.0\%. There is a small increase in the
uncertainty for distributions where there is at least one
jet in the endcap region of the detector.

\subsection{Jet energy resolution}

The JER is measured in data using the \pt balance in dijet events
\cite{jeres}. Based on these measurements, the
resolution effects are corrected using simulated events. To
study the effect of the difference between the simulated and the measured
resolution, several sets of unfolded distributions
are obtained using response matrices from the default resolution matrix and
changing the jet resolution within its estimated uncertainty. Alternatively,
the response matrix is constructed by convolving the generator level
distribution with the measured resolution. The measured distribution is unfolded
by this response matrix vis-a-vis the response matrix determined using
fully simulated sample of \PYTHIA{6} events. These two estimates
provide independent descriptions of the detector modeling and the difference
is used as a measure of the systematic uncertainty due to detector
performance. Position resolution affects the measurement of the jet direction,
and it is estimated using simulated multijet events and validated with data.

Uncertainties owing to the JER are found to be between
0.1--10\% in the three-jet mass, 0.3--15\% in the four-jet mass,
0.1--10\% in the scaled jet energies and 0.2--8.2\% in the angular variables.

\subsection{Model dependence in unfolding}

Unfolded distributions are obtained using two different response matrices
derived from \PYTHIA{6} and from \MADGRAPHPYTHIASIX
simulations. The difference in the unfolded values, due to the choice
of response functions, gives a measure of the systematic uncertainty.
The uncertainties are at the level of 0.1--6.0\% in the three-jet mass,
0.1--3.0\% in the scaled energy distributions of the three-jet variables,
0.1--8.0\% in the four-jet mass and 0.1--6.2\% in the angular variables in
the four-jet samples. The uncertainties in the scaled jet energy increases
by a few percent for the samples with lower values of leading jet \pt.

Unfolding has been carried out using \PYTHIA{6} and \MADGRAPHPYTHIASIX
samples, which has the same hadronization model.
To test the effects of different hadronization models, MC samples from
\HERWIGpp, which provides a different PS and hadronization approach,
are used. However, the simulated event sample generated using
\HERWIGpp is statistically
inadequate to be used in a complete unfolding procedure. The difference
between bin-by-bin correction factors obtained with \PYTHIA{6} and
\HERWIGpp is found to be somewhat larger than the uncertainty due
to the difference in the unfolding matrices: 0.1--12\% in the scaled energy
distributions of three-jet variables, 0.1--7.7\% in the angular variables in
the four-jet samples and 0.1--11.6\% in the jet masses.
The larger values from the two estimates are chosen as the
systematic uncertainties due to unfolding.

\subsection{Event selection}

Jet candidates are required to pass certain criteria \cite{jetid} designed to
reduce unwanted detector effects. This analysis uses jets identified with
very restrictive criteria on the ratio of the energy carried by neutral to
that carried by charged particles. The effect of using these criteria is
tested by reevaluating the same distributions with jets selected after
relaxing the selection on the fractions of the energy carried by the neutral
and the charged particles. Also, the selection on $\ETm$
is changed, and the effect of this is estimated from the difference in
the observed distributions. The uncertainty due to the event selection is found
to be below 0.2\%.

\subsection{Overall uncertainty}

The first three sources mentioned above are the dominant sources of
systematic uncertainty. The contributions to the uncertainty from the selection
requirements and pileup effects are found to be negligible.
The uncertainties are calculated for each bin of the measured distributions
and are added in quadrature. The overall systematic uncertainty is found to
be smaller than the statistical uncertainty for most of the bins. Typical
uncertainties for the six variables studied in this analysis are summarized
in Table \ref{tab:err}.

\begin{table*}[htbp]
\topcaption{Uncertainty ranges among the different bins in the topological
           distributions of the three- and four-jet variables.}
\centering\begin{tabular}{lcc}\hline
\multicolumn{1}{c}{Uncertainty source} &
\multicolumn{2}{c}{Uncertainty (\%) for leading jet \pt} \\ \cline{2-3}
                   & 190--300\GeV
                   & $>$500\GeV  \\ \hline
\multicolumn{3}{c}{Three-jet mass} \\ \hline
Jet and event selection       & 0.1        & 0.1        \\
Jet energy scale              & 0.3--5.0  & 0.2--5.5  \\
Jet resolution                & 0.1--10.0 & 0.2--6.0  \\
Model dependence in unfolding & 0.2--11.0 & 0.2--5.0  \\ \hline
Total systematic uncertainty  & 0.3--12.7 & 0.2--7.9  \\
Statistical uncertainty       & 1.4--14.5 & 0.7--10.2 \\ \hline
\multicolumn{3}{c}{Scaled energy of the leading jet} \\ \hline
Jet and event selection       & 0.1        & 0.1        \\
Jet energy scale              & 0.1--1.9  & 0.1--1.4  \\
Jet resolution                & 0.2--6.2  & 0.1--5.4  \\
Model dependence in unfolding & 0.1--6.0  & 0.5--3.6  \\ \hline
Total systematic uncertainty  & 0.8--7.2  & 1.1--5.6  \\
Statistical uncertainty       & 1.6--17.2 & 0.6--14.2 \\ \hline
\multicolumn{3}{c}{Scaled energy of the second-leading jet} \\ \hline
Jet and event selection       & 0.1        & 0.1        \\
Jet energy scale              & 0.1--2.0  & 0.1--2.0  \\
Jet resolution                & 0.1--5.0  & 0.1--4.2  \\
Model dependence in unfolding & 0.4--9.0  & 0.1--3.5  \\ \hline
Total systematic uncertainty  & 1.0--8.3  & 0.1--4.6  \\
Statistical uncertainty       & 1.3--16.4 & 0.9--8.0  \\ \hline
\multicolumn{3}{c}{Four-jet mass} \\ \hline
Jet and event selection       & 0.1        & 0.1        \\
Jet energy scale              & 0.4--6.9  & 0.3--7.0  \\
Jet resolution                & 0.4--11.7 & 0.2--4.9  \\
Model dependence in unfolding & 0.3--7.0  & 0.5--8.1  \\ \hline
Total systematic uncertainty  & 0.4--13.7 & 0.5--11.6 \\
Statistical uncertainty       & 3.1--30.9 & 1.4--12.5 \\ \hline
\multicolumn{3}{c}{Bengtsson--Zerwas angle} \\ \hline
Jet and event selection       & 0.1        & 0.1        \\
Jet energy scale              & 0.1--3.0  & 0.2--2.4  \\
Jet resolution                & 0.4--5.4  & 0.2--5.0  \\
Model dependence in unfolding & 0.3--3.5  & 0.1--6.4  \\ \hline
Total systematic uncertainty  & 1.4--5.9  & 1.0--8.1  \\
Statistical uncertainty       & 5.1--8.4  & 2.8--4.0  \\ \hline
\multicolumn{3}{c}{Nachtmann--Reiter angle} \\ \hline
Jet and event selection       & 0.1        & 0.1        \\
Jet energy scale              & 0.1--1.0  & 0.1--1.1  \\
Jet resolution                & 0.1--4.6  & 0.2--2.1  \\
Model dependence in unfolding & 0.2--2.1  & 0.4--5.0  \\ \hline
Total systematic uncertainty  & 0.9--5.0  & 0.9--5.2  \\
Statistical uncertainty       & 3.4--4.2  & 1.3--1.6  \\ \hline
\end{tabular}
\label{tab:err}
\end{table*}

\section{Results}\label{sec:results}

\subsection{Comparison with models}

The normalized differential distributions, corrected for detector effects, are
plotted as a function of the three- and four-jet inclusive variables and
compared with predictions from the four MC
models: \PYTHIA{6}, \PYTHIA{8}, \MADGRAPHPYTHIASIX
and \HERWIGpp. The variables considered
for these comparisons are three-jet mass, scaled energies of the leading
and next-to-leading jet in the three-jet sample in the three-jet CM
frame, four-jet mass, and the two angles $\cbz$ and $\cnr$.

\begin{figure*}[h!tbp]
\centering
 \includegraphics[width=\cmsFigWidth]{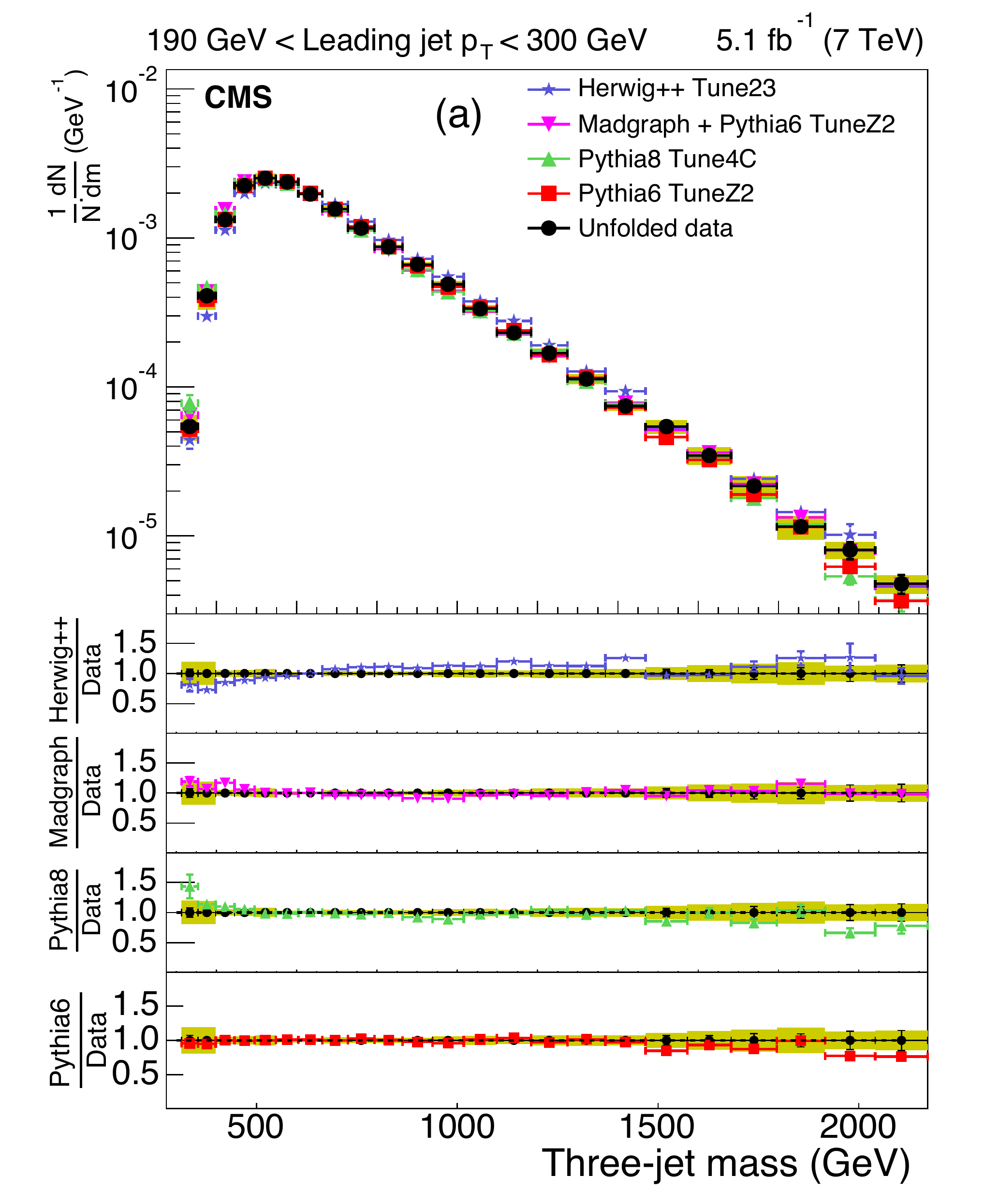}
 \includegraphics[width=\cmsFigWidth]{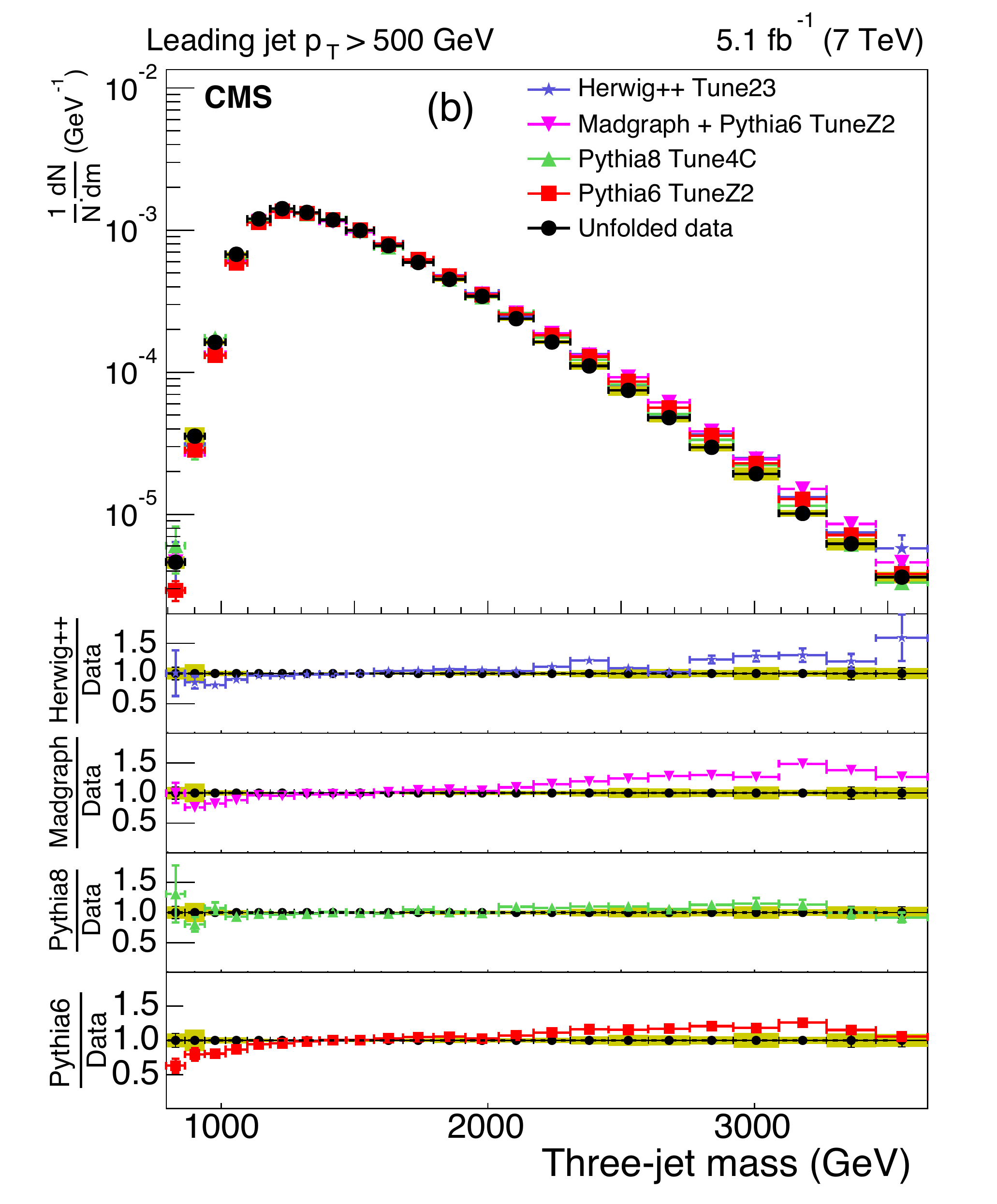}

\caption{Distribution of the three-jet mass superposed with predictions from
         four MC models: \PYTHIA{6}, \PYTHIA{8},
         \MADGRAPHPYTHIASIX, \HERWIGpp. The
         distributions are obtained from inclusive three-jet sample with the
         jets restricted in the $\abs{y}$ region $0.0 <\abs{y}< 2.5$, and with
         leading-jet \pt between 190 and 300\GeV (a) or above 500\GeV (b).
         The data points are shown with statistical uncertainty only and the
         bands indicate the statistical and systematic uncertainties combined
         in quadrature. The lower panels of each plot show the ratios of MC
         predictions to the data. The ratios are shown with statistical
         uncertainty in the data as well as in the MC, while the band
         shows combined statistical and systematic uncertainties.}
\label{fig:3jetmass}
\end{figure*}

\begin{figure}[h!tbp]
\centering
 \includegraphics[width=\cmsFigWidth]{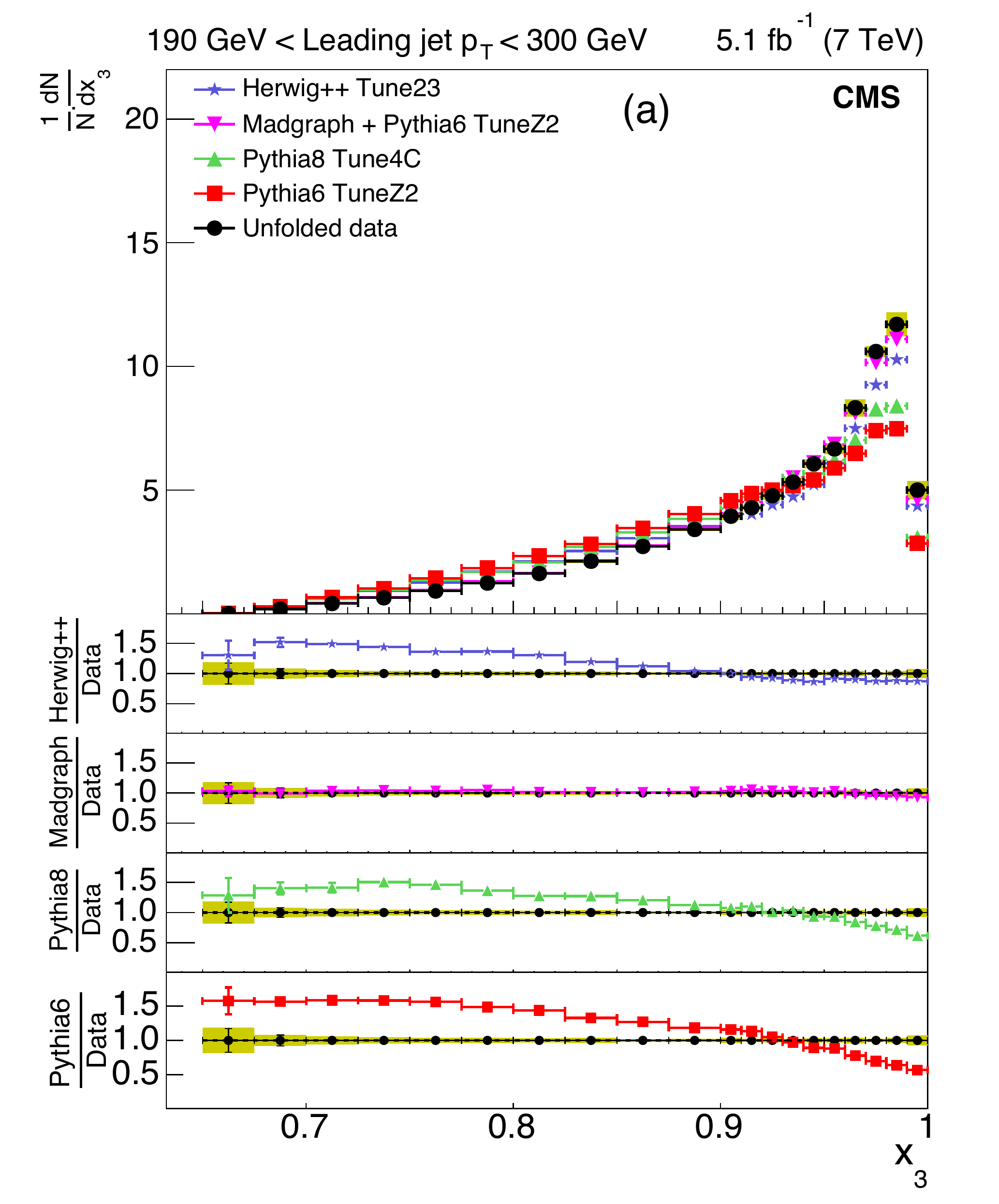}
 \includegraphics[width=\cmsFigWidth]{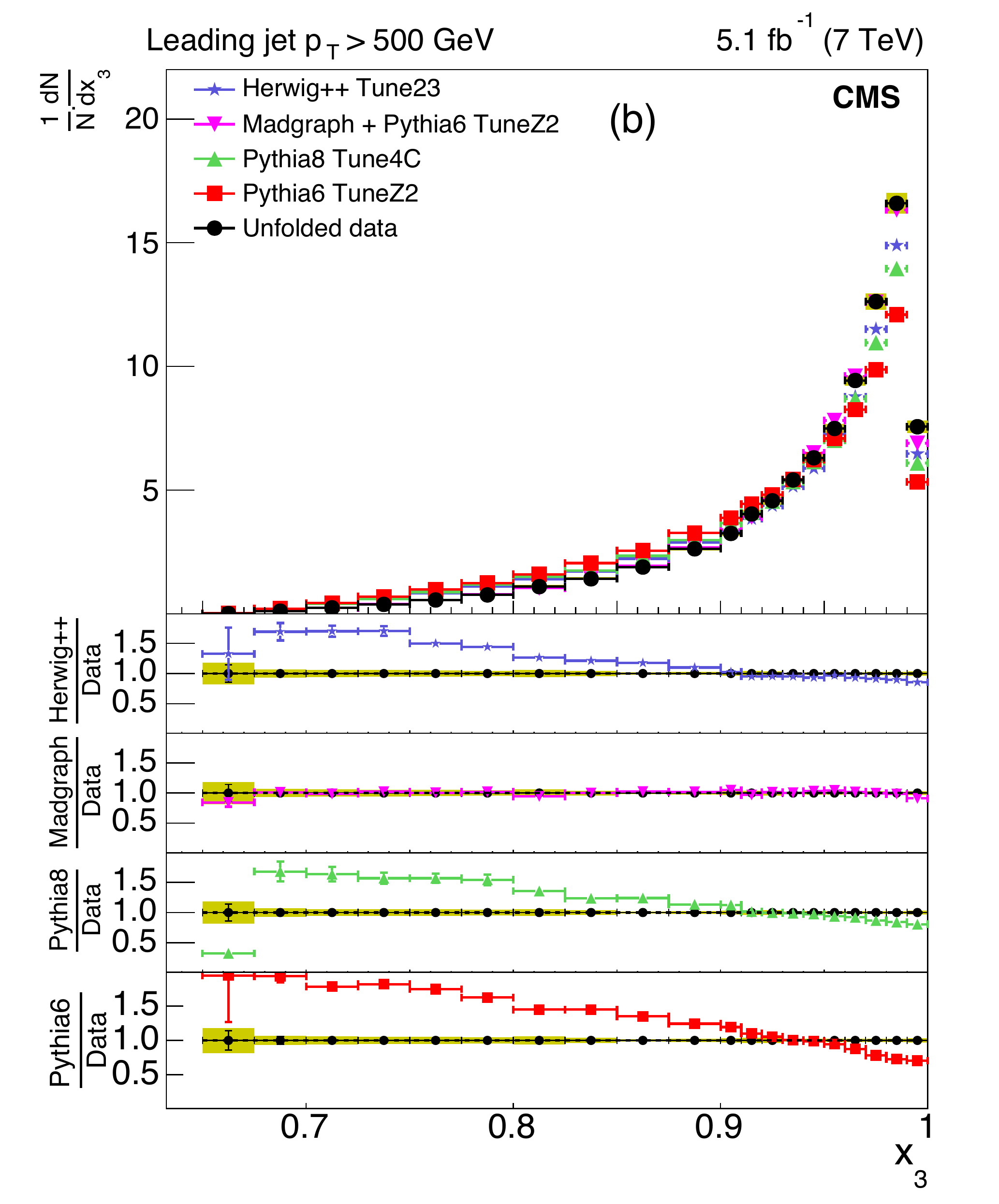}

\caption{Corrected normalized distribution of scaled energy of the leading-jet
         in the inclusive three-jet sample. The other explanations are the same
         as Fig.~\ref{fig:3jetmass}.}
\label{fig:3jetx0}
\end{figure}

\begin{figure}[h!tbp]
\centering
 \includegraphics[width=\cmsFigWidth]{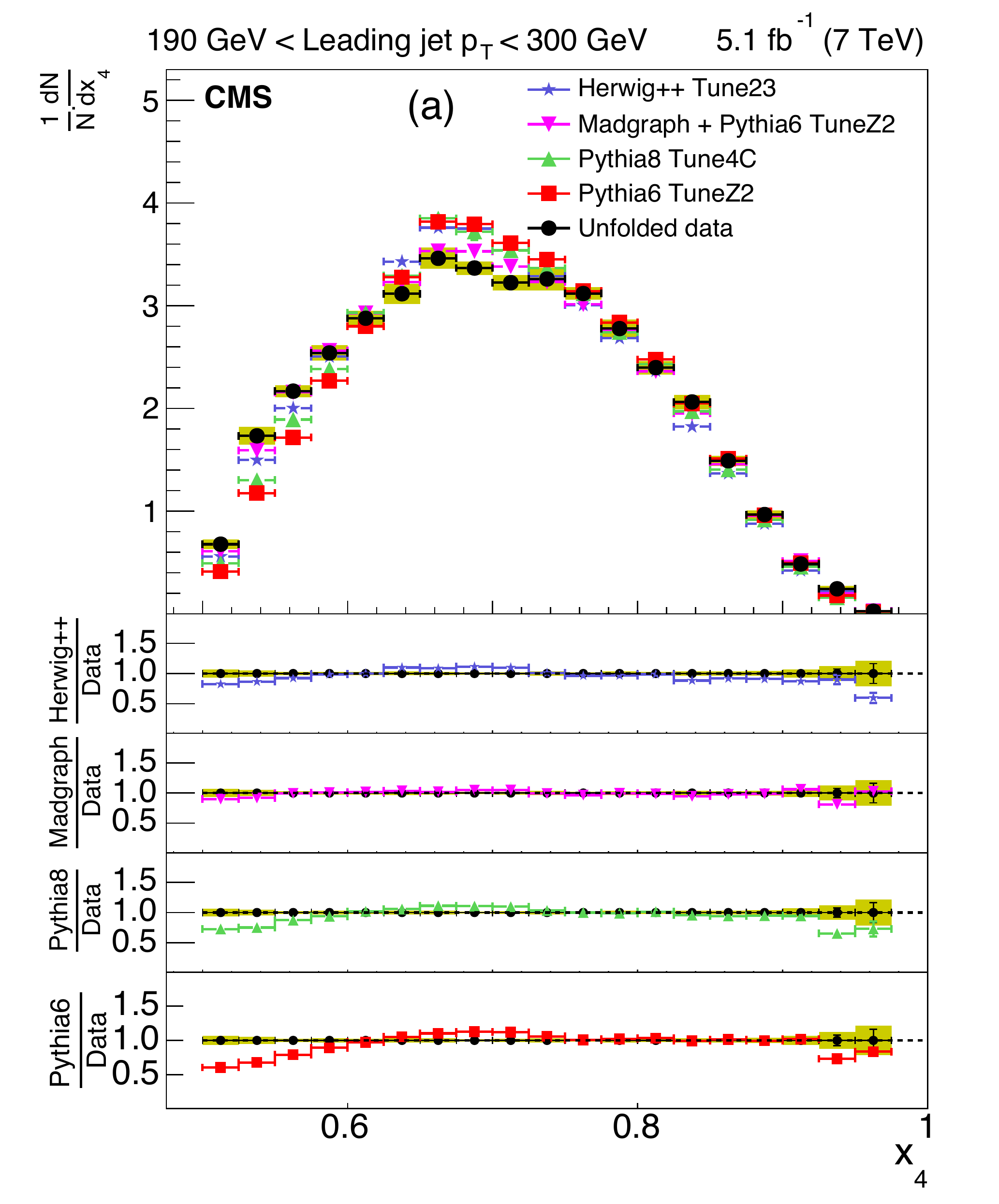}
 \includegraphics[width=\cmsFigWidth]{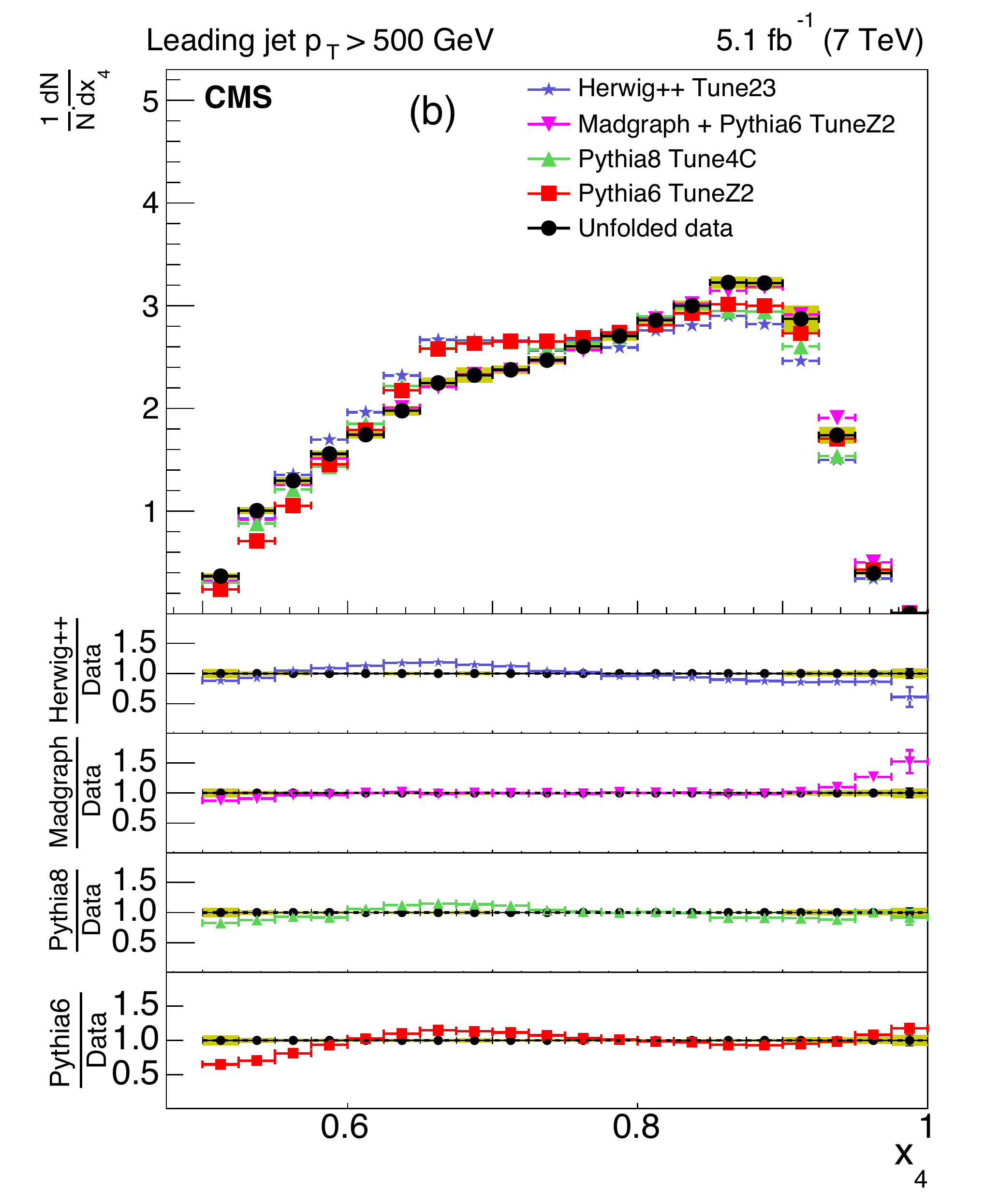}

\caption{Corrected normalized distribution of scaled energy of the
         second-leading jet in the inclusive three-jet sample. The other
         explanations are the same as Fig.~\ref{fig:3jetmass}.}
\label{fig:3jetx1}
\end{figure}

\begin{figure}[h!tbp]
\centering
 \includegraphics[width=\cmsFigWidth]{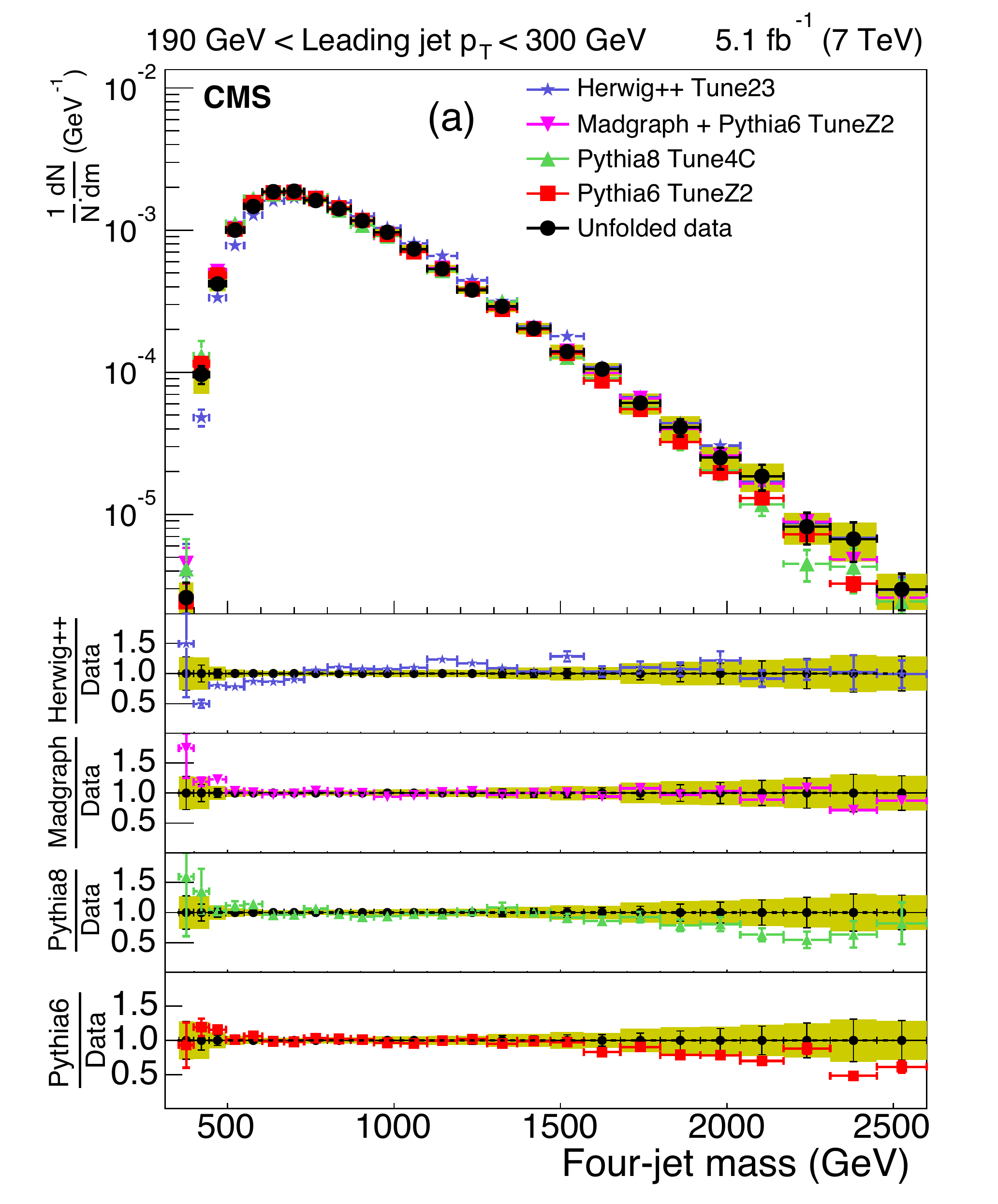}
 \includegraphics[width=\cmsFigWidth]{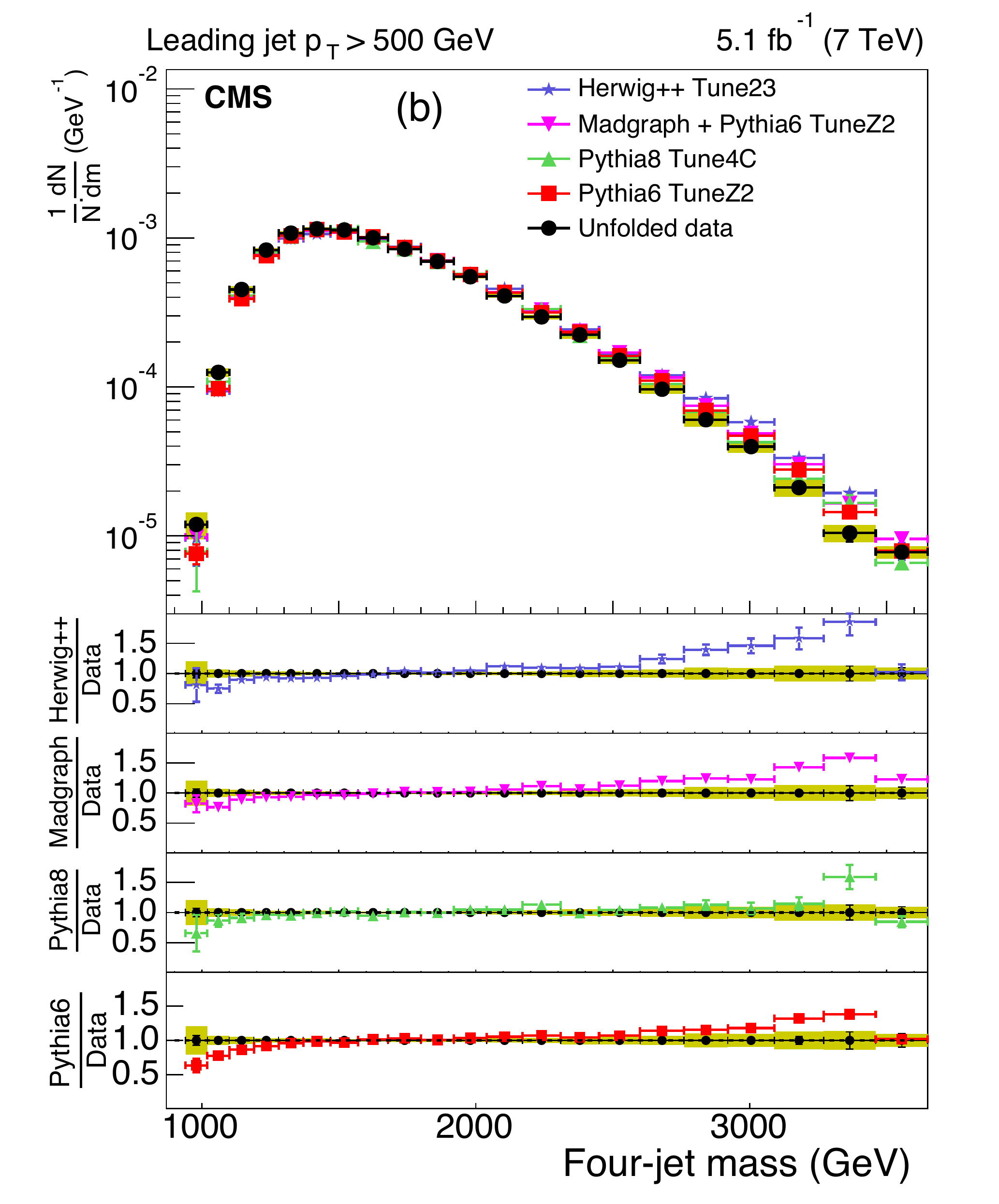}

\caption{Corrected normalized distribution of four-jet mass. The other
         explanations are the same as Fig.~\ref{fig:3jetmass}.}
\label{fig:4jetmass}
\end{figure}

\begin{figure}[h!tbp]
\centering
 \includegraphics[width=\cmsFigWidth]{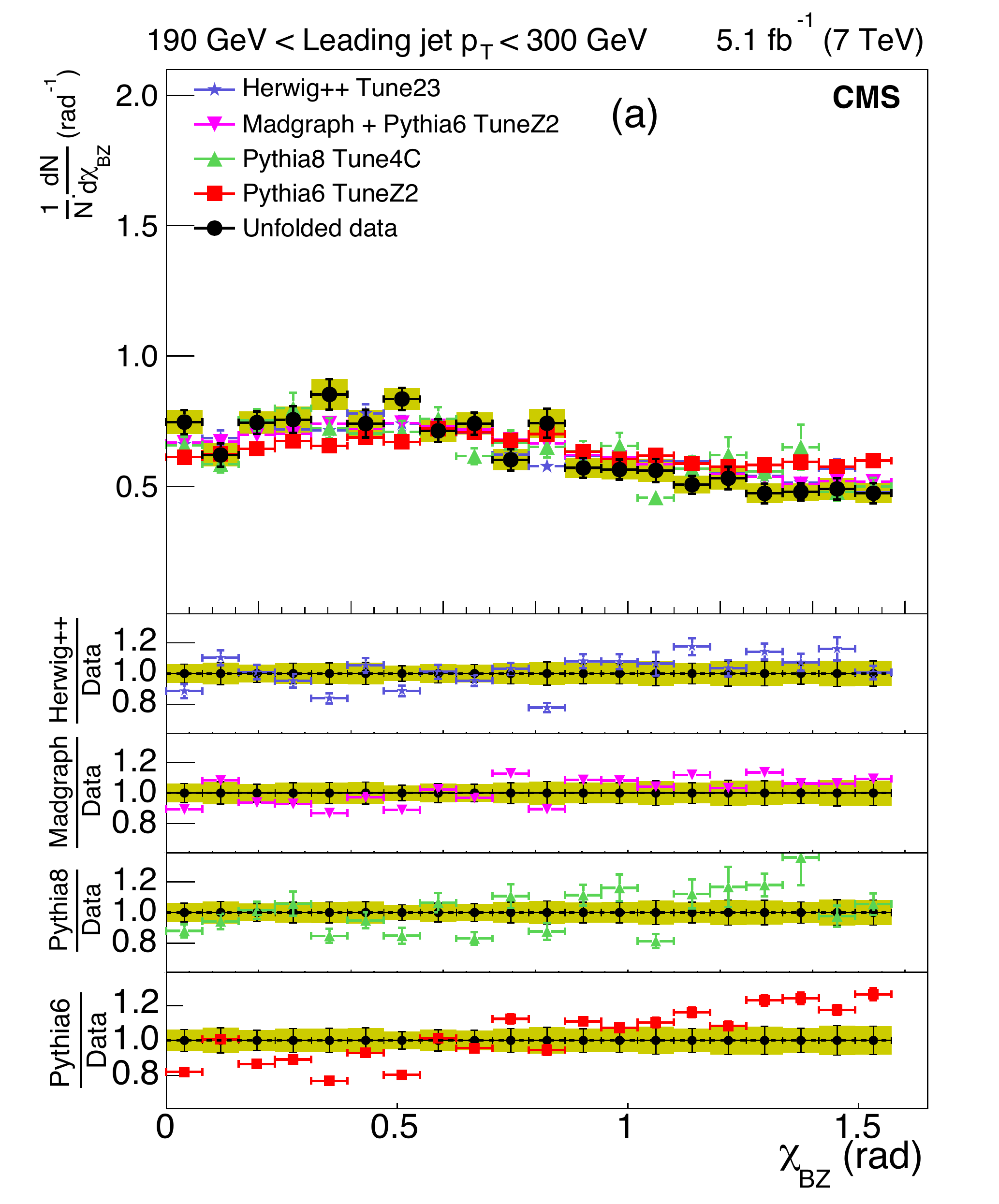}
 \includegraphics[width=\cmsFigWidth]{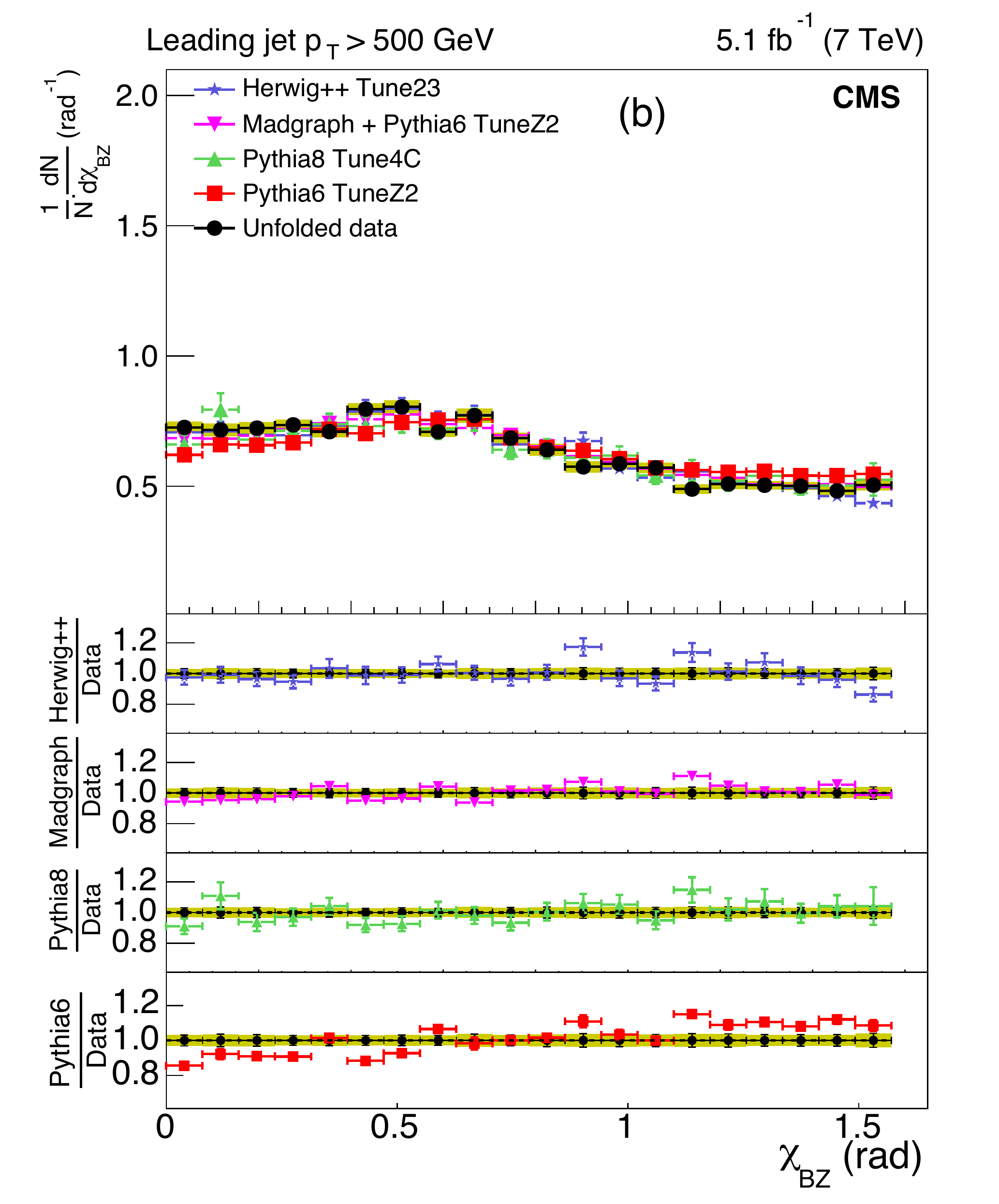}

\caption{Corrected normalized distribution of the Bengtsson--Zerwas angle.
         The other explanations are the same as Fig.~\ref{fig:3jetmass}.}
\label{fig:thetaBZ}
\end{figure}

For the comparison plots (Figs. \ref{fig:3jetmass}--\ref{fig:thetaBZ}),
the upper panel shows the data and the model
predictions with the corresponding statistical uncertainty. For the data,
the shaded area shows the statistical and systematic uncertainties added in
quadrature. The lower panels in each plot show the ratio of MC
prediction to the data for each model. Comparisons are made for two different
ranges of the leading jet \pt: $190 < \pt < 300\GeV$ and $\pt >500$\GeV.

Figure \ref{fig:3jetmass} shows the normalized corrected differential
distribution as a function of the three-jet mass for two ranges of the
leading-jet \pt.
The three-jet mass distribution broadens for larger \pt thresholds.
The models show varying degrees of success for the different ranges of
leading-jet \pt. Most models differ from the data in the low-mass
spectrum. The \PYTHIA{6} simulation provides a good description
of the data in the lower \pt bin, while it has a larger deviation in the
higher \pt bin. The mean difference is at the level of 1.8--4.0\%.
Predictions from  \MADGRAPHPYTHIASIX and \PYTHIA{8}
agree with the data to within 4.5\%. \HERWIGpp provides
the worst agreement among the four models -- the mean difference is at the
level of 4.0--15\%.

Figure \ref{fig:3jetx0} shows the corrected normalized differential distribution
as a function of the scaled leading-jet energy in the inclusive three-jet
sample. The distributions peak close to 1 and the peaks get sharper for
higher leading-jet \pt range. The scaled leading-jet energy $x_3$
is expected to follow a linear rise from $\frac{2}{3}$ to 1 for a
phase space model, which has only energy-momentum conservation,
while QCD predicts a deviation from linearity at higher values of $x_3$.
This feature is observed in the data, particularly for higher \pt bins.
Only \MADGRAPHPYTHIASIX provides a consistent
description of the data. The agreement improves for the sample with leading-jet
\pt above 500\GeV. The difference between the predictions from
\MADGRAPHPYTHIASIX and the data are at the level of 3.5--6.1\%.

Figure \ref{fig:3jetx1} shows the corrected normalized differential distribution
as a function of the scaled energy of the second-leading jet, $x_4$, in the
inclusive three-jet sample. For kinematic reasons, $x_4$ is expected to lie
between 1/2 and 1. The distribution peaks around 0.65 for the low
\pt threshold sample. The peak shifts to higher values of $x_4$ and the
distribution becomes broader for the larger \pt threshold sample.
Predictions from \MADGRAPHPYTHIASIX agree with data
to within 3.1\%. Predictions from \PYTHIA{6} as well as
\PYTHIA{8} deviate by as much as 10\% or more from the data.
Predictions from \HERWIGpp also shows a large deviation at higher \pt bins.

Figure \ref{fig:4jetmass} shows comparisons of the corrected normalized
differential distribution as a function of the four-jet mass for
the four MC models. The distribution broadens at higher minimum \pt value.
As can be seen from the figure, \HERWIGpp provides the worst
comparison. The average deviations are at the level of 15\%
for many of the distributions, particularly for the sample with leading-jet
\pt between 190 and 300\GeV.
The level of agreement for the other three MC models is better than 10\%
over the entire \pt region.

The sub-leading jets in the four-jet event category are predominantly due to
the secondary splitting of partons. In case of gluon splitting, they
can be due to a $\PQq\PAQq$ pair or gluons. Both the angular
distributions, $\cnr$ and $\cbz$, are different for these two scenarios and
are representative of the colour factors for these couplings.

Figure \ref{fig:thetaBZ} shows similar comparisons for the
Bengtsson--Zerwas angle. Because the azimuthal angle is not defined for the
back-to-back jets, the opening angle between the two leading and two
nonleading jets is required to be less than 160$^{\circ}$.
As can be seen from the average deviation of the ratios from unity,
predictions from \MADGRAPHPYTHIASIX and \HERWIGpp represent
the data well, while those from \PYTHIA{6} do poorly.

\begin{figure}[h!tbp]
\centering
 \includegraphics[width=\cmsFigWidth]{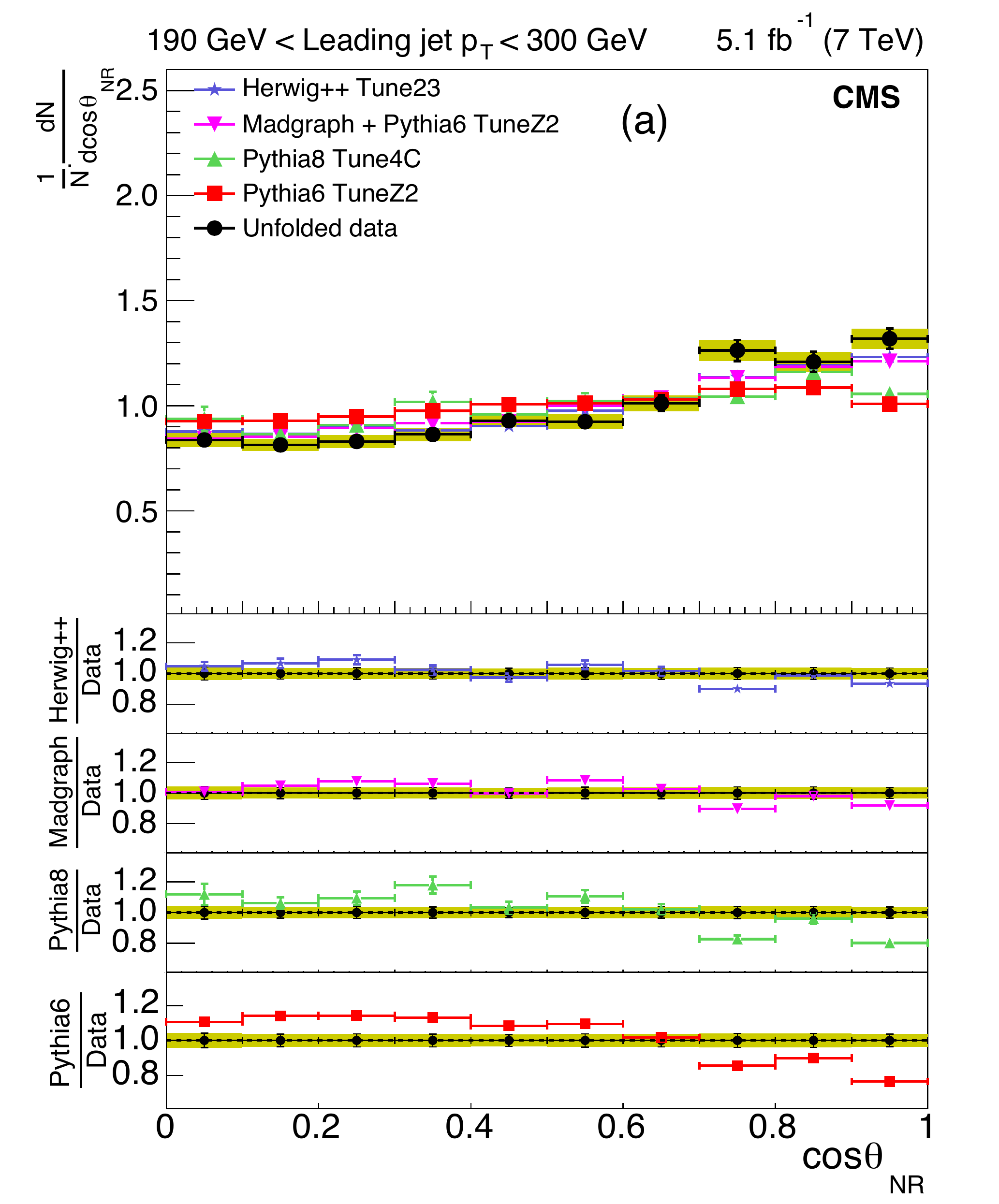}
 \includegraphics[width=\cmsFigWidth]{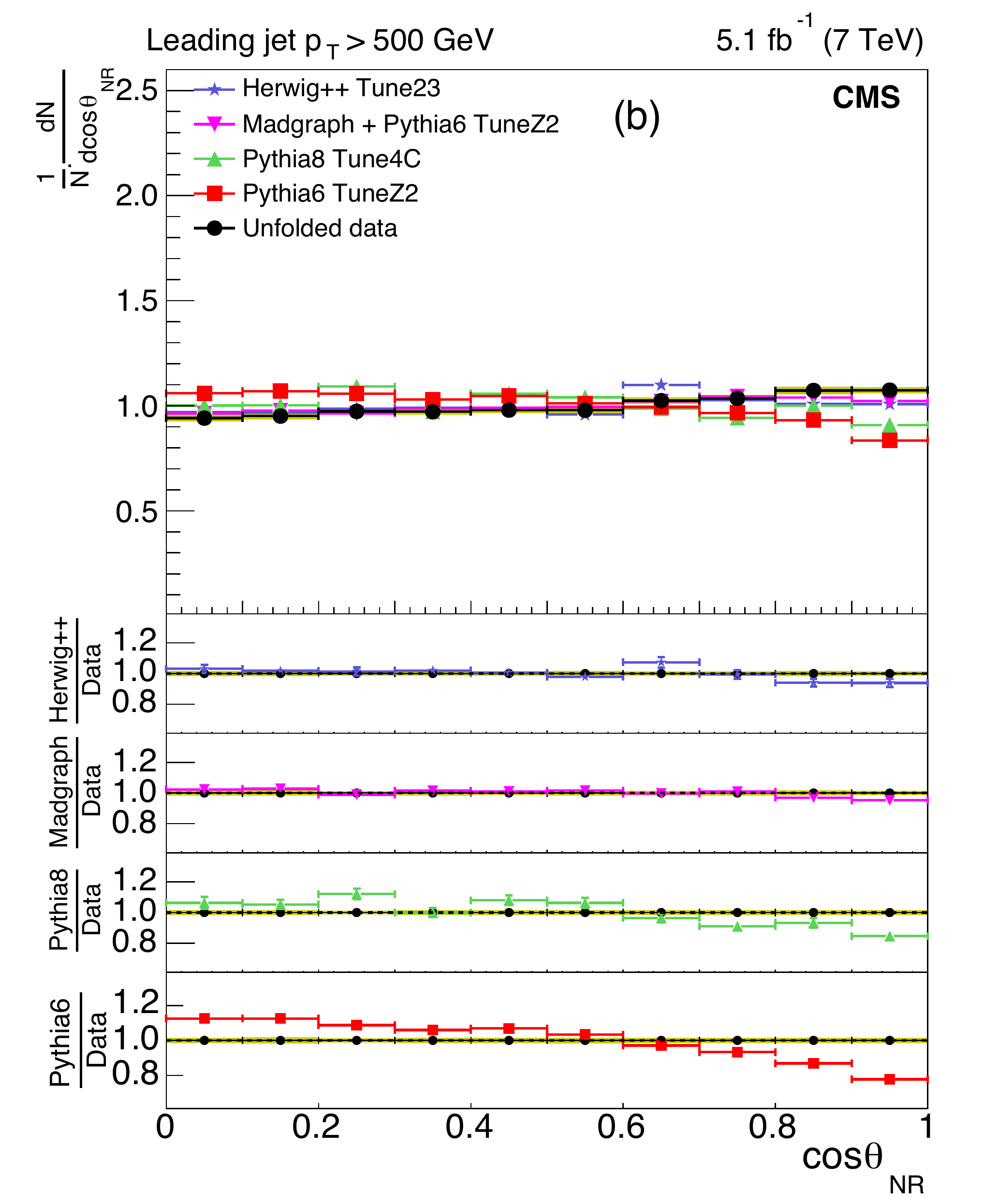}

\caption{Corrected normalized distribution of the cosine of the
         Nachtmann--Reiter angle. The other explanations are the same as
         Fig.~\ref{fig:3jetmass}.}
\label{fig:cthetaNR}
\end{figure}

Figure \ref{fig:cthetaNR} shows the corrected normalized differential
distribution as a function of the cosine of the Nachtmann--Reiter angle in the
inclusive four-jet sample. Most of the models follow the broad features of
the data. However, the degree of agreement with data is different among
models. \MADGRAPHPYTHIASIX provides the best description of
the data; \HERWIGpp with angular ordering in the parton shower is close
to the data (the agreement is better than 5\%), while \PYTHIA{6}
has the largest deviation (the agreement is typically between 10--12\%).

\subsection{Effect of hadronization, underlying event, and PDFs}

The disagreement between data and the MC models may arise from
the implementation of nonperturbative components in the simulation
due to the fragmentation model or the choice of PDF
set. These effects have been investigated by studying the
uncertainties due to hadronization model and PDF parametrization.

The MC models have different ways of modeling the
underlying events and hadronization of the partons into hadrons. This may
result in different predictions of the distributions of multijet variables
depending on whether they are computed at the hadron or at the parton level.
This effect has been investigated by studying two different MC models:
\PYTHIA{6} and \HERWIGpp. This is done by evaluating
the distributions at the parton and hadron level. \PYTHIA{6} uses
the \textsc{Lund} string model, while \HERWIGpp uses the cluster
model. Also, colour reconnections are done differently in the two models.
A generator-level study is carried out for both these models, where
the effect of hadronization is studied using distributions from jets at
parton- and hadron-level. The ratio of the parton- to the
hadron-level distribution is then compared. The mean difference between
the two hadronization models is typically less than 5\%.

Comparisons are also made to different tunes of the underlying event models
within \PYTHIA{6}. The tunes (D6T, DW, P0, Z1, Z2, Z2*)
\cite{z2tune,z1tune,d6t,p0,z2star}
differ in the cutoff used to regularize the 1/$\pt^4$ divergence
for final-state partons, the ordering of the showers (virtuality ordering
\vs \pt ordering), multiparton interaction model, PDFs, and
data sets used in the tune. The resulting distributions agree
typically within 5\%, so the disagreements with the data cannot be
fully explained by this effect.

{\tolerance=800
The MC models use CTEQ6 as the default PDF parametrization.
There are many different PDF sets, which are based on different input data,
assumptions, and parametrizations. Thus any calculation of a cross section or
distributions in the simulation depends on the choice of PDF set. Also, each
PDF set has its own errors from its parametric
assumptions and data input to fitting. The effect of the PDF set choice on the
multijet variables is calculated according to the recommendation of PDF4LHC
group \cite{pdf4lhc,pdf4lhc2}. Since comparisons are made only with leading
order Monte Carlo models in this paper, only two leading order PDF sets are used
in this comparison: CTEQ6l and MSTW2008lo68cl~\cite{mstw}. The uncertainties are
found to be typically at the level of 1.0--2.0\% depending on the variable type
and \pt range considered.
\par}

\section{Summary}\label{sec:summ}

{\tolerance=800
Distributions of topological variables for inclusive three- and four-jet events
in pp collisions measured with the CMS detector at a centre-of-mass energy of
7\TeV  were presented using a data sample corresponding to an integrated
luminosity of 5.1\fbinv. The distributions were corrected for detector effects,
and systematic uncertainties were estimated. These corrected distributions were
compared with the predictions from four LO MC models: \PYTHIA{6},
\PYTHIA{8}, \HERWIGpp, and \MADGRAPHPYTHIASIX.
\par}

Distributions of three- and four-jet invariant mass from all models show
significant deviation from the data at high mass. The fact that all models
have a common PDF suggests that the PDF errors at high mass are
underestimated. The PDFs at high invariant mass have recently been
constrained by CMS using dijet \pt distributions\cite{smp12-028}.

The \MADGRAPH simulations are based on tree-level
calculations for two-, three-, and four-parton final states, while
\PYTHIA and \HERWIGpp can have only two partons in the final
state before showering. Not surprisingly, the three-jet predictions of
\MADGRAPHPYTHIASIX give a more consistent description of the
distributions studied in this analysis. The notable exception is at high
$x_4$ (the next-to-leading jet), where two jets carry most of the CM energy.
The difference is probably due to a double counting of three-parton with
two-parton (with a parton from showering) final states.

The \PYTHIA and \HERWIGpp models give poor descriptions of
the energy fractions in the three-jet final state. In particular, the
distributions of $x_3$ (the leading jet) show large shape differences
between data and theory that are inconsistent with PDFs or hadronization
model uncertainties. Since the distributions from \MADGRAPHPYTHIASIX
agree with those from the data, the discrepancies with \PYTHIA and
\HERWIGpp are likely due to missing higher multiplicity ME, which are
present in \MADGRAPH.

All the models compared in this study do remarkably well describing the
four-jet Bengtsson--Zerwas angle. The \PYTHIA models
have some systematic deviation from the data in describing the Nachtmann--Reiter
angle. Parton showers with angular ordering, as implemented in
\HERWIGpp, yield a better agreement with the measured data for these
angular variables.

\begin{acknowledgments}
\hyphenation{Bundes-ministerium Forschungs-gemeinschaft Forschungs-zentren} We congratulate our colleagues in the CERN accelerator departments for the excellent performance of the LHC and thank the technical and administrative staffs at CERN and at other CMS institutes for their contributions to the success of the CMS effort. In addition, we gratefully acknowledge the computing centres and personnel of the Worldwide LHC Computing Grid for delivering so effectively the computing infrastructure essential to our analyses. Finally, we acknowledge the enduring support for the construction and operation of the LHC and the CMS detector provided by the following funding agencies: the Austrian Federal Ministry of Science, Research and Economy and the Austrian Science Fund; the Belgian Fonds de la Recherche Scientifique, and Fonds voor Wetenschappelijk Onderzoek; the Brazilian Funding Agencies (CNPq, CAPES, FAPERJ, and FAPESP); the Bulgarian Ministry of Education and Science; CERN; the Chinese Academy of Sciences, Ministry of Science and Technology, and National Natural Science Foundation of China; the Colombian Funding Agency (COLCIENCIAS); the Croatian Ministry of Science, Education and Sport, and the Croatian Science Foundation; the Research Promotion Foundation, Cyprus; the Ministry of Education and Research, Estonian Research Council via IUT23-4 and IUT23-6 and European Regional Development Fund, Estonia; the Academy of Finland, Finnish Ministry of Education and Culture, and Helsinki Institute of Physics; the Institut National de Physique Nucl\'eaire et de Physique des Particules~/~CNRS, and Commissariat \`a l'\'Energie Atomique et aux \'Energies Alternatives~/~CEA, France; the Bundesministerium f\"ur Bildung und Forschung, Deutsche Forschungsgemeinschaft, and Helmholtz-Gemeinschaft Deutscher Forschungszentren, Germany; the General Secretariat for Research and Technology, Greece; the National Scientific Research Foundation, and National Innovation Office, Hungary; the Department of Atomic Energy and the Department of Science and Technology, India; the Institute for Studies in Theoretical Physics and Mathematics, Iran; the Science Foundation, Ireland; the Istituto Nazionale di Fisica Nucleare, Italy; the Ministry of Science, ICT and Future Planning, and National Research Foundation (NRF), Republic of Korea; the Lithuanian Academy of Sciences; the Ministry of Education, and University of Malaya (Malaysia); the Mexican Funding Agencies (CINVESTAV, CONACYT, SEP, and UASLP-FAI); the Ministry of Business, Innovation and Employment, New Zealand; the Pakistan Atomic Energy Commission; the Ministry of Science and Higher Education and the National Science Centre, Poland; the Funda\c{c}\~ao para a Ci\^encia e a Tecnologia, Portugal; JINR, Dubna; the Ministry of Education and Science of the Russian Federation, the Federal Agency of Atomic Energy of the Russian Federation, Russian Academy of Sciences, and the Russian Foundation for Basic Research; the Ministry of Education, Science and Technological Development of Serbia; the Secretar\'{\i}a de Estado de Investigaci\'on, Desarrollo e Innovaci\'on and Programa Consolider-Ingenio 2010, Spain; the Swiss Funding Agencies (ETH Board, ETH Zurich, PSI, SNF, UniZH, Canton Zurich, and SER); the Ministry of Science and Technology, Taipei; the Thailand Center of Excellence in Physics, the Institute for the Promotion of Teaching Science and Technology of Thailand, Special Task Force for Activating Research and the National Science and Technology Development Agency of Thailand; the Scientific and Technical Research Council of Turkey, and Turkish Atomic Energy Authority; the National Academy of Sciences of Ukraine, and State Fund for Fundamental Researches, Ukraine; the Science and Technology Facilities Council, UK; the US Department of Energy, and the US National Science Foundation.

Individuals have received support from the Marie-Curie programme and the European Research Council and EPLANET (European Union); the Leventis Foundation; the A. P. Sloan Foundation; the Alexander von Humboldt Foundation; the Belgian Federal Science Policy Office; the Fonds pour la Formation \`a la Recherche dans l'Industrie et dans l'Agriculture (FRIA-Belgium); the Agentschap voor Innovatie door Wetenschap en Technologie (IWT-Belgium); the Ministry of Education, Youth and Sports (MEYS) of the Czech Republic; the Council of Science and Industrial Research, India; the HOMING PLUS programme of Foundation for Polish Science, cofinanced from European Union, Regional Development Fund; the Compagnia di San Paolo (Torino); the Consorzio per la Fisica (Trieste); MIUR project 20108T4XTM (Italy); the Thalis and Aristeia programmes cofinanced by EU-ESF and the Greek NSRF; and the National Priorities Research Program by Qatar National Research Fund.
\end{acknowledgments}

\bibliography{auto_generated}

\cleardoublepage \appendix\section{The CMS Collaboration \label{app:collab}}\begin{sloppypar}\hyphenpenalty=5000\widowpenalty=500\clubpenalty=5000\textbf{Yerevan Physics Institute,  Yerevan,  Armenia}\\*[0pt]
V.~Khachatryan, A.M.~Sirunyan, A.~Tumasyan
\vskip\cmsinstskip
\textbf{Institut f\"{u}r Hochenergiephysik der OeAW,  Wien,  Austria}\\*[0pt]
W.~Adam, T.~Bergauer, M.~Dragicevic, J.~Er\"{o}, M.~Friedl, R.~Fr\"{u}hwirth\cmsAuthorMark{1}, V.M.~Ghete, C.~Hartl, N.~H\"{o}rmann, J.~Hrubec, M.~Jeitler\cmsAuthorMark{1}, W.~Kiesenhofer, V.~Kn\"{u}nz, M.~Krammer\cmsAuthorMark{1}, I.~Kr\"{a}tschmer, D.~Liko, I.~Mikulec, D.~Rabady\cmsAuthorMark{2}, B.~Rahbaran, H.~Rohringer, R.~Sch\"{o}fbeck, J.~Strauss, W.~Treberer-Treberspurg, W.~Waltenberger, C.-E.~Wulz\cmsAuthorMark{1}
% ...
\vskip\cmsinstskip
% ...
1:~~Also at Vienna University of Technology, Vienna, Austria\\
2:~~Also at CERN, European Organization for Nuclear Research, Geneva, Switzerland\\
% ...

\end{sloppypar}
\end{document}